\documentclass[11pt]{article}
\usepackage{amsmath}
\usepackage{amsfonts}
\usepackage{amssymb}
\usepackage{graphicx}
\usepackage{bm}
\usepackage{hyperref}
\usepackage{color}
\usepackage{mathrsfs}
\usepackage{epsf}

\begin{document}
\newcommand{\la}{\label}
\newcommand{\ba}{\begin{align}}
\newcommand{\h}{{(h)}}
\newcommand{\al}{{(\alpha)}}
\newcommand{\ls}{\!\prec\!\!}
\newcommand{\rs}{\!\!\succ\;}
\newcommand{\nol}{:\!}
\newcommand{\nor}{\!:}
\newcommand{\w}{\widetilde}
\newcommand{\der}{\partial}
\title{Critical curves in conformally invariant statistical systems}
\author{I. Rushkin, E. Bettelheim, I. A. Gruzberg, P. Wiegmann
\\ \\
The James Franck Institute, The University of Chicago\\
5640 S. Ellis Avenue, Chicago, Il 60637 USA }

\date{January 17, 2007}

\maketitle

\begin{abstract}

We consider critical curves --- conformally invariant curves that
appear at critical points of two-dimensional statistical mechanical
systems. We show how to describe these curves in terms of the
Coulomb gas formalism of conformal field theory (CFT). We also
provide links between this description and the stochastic (Schramm-)
Loewner evolution (SLE). The connection appears in the long-time
limit of stochastic evolution of various SLE observables related to
CFT primary fields. We show how the multifractal spectrum of
harmonic measure and other fractal characteristics of critical
curves can be obtained.

\end{abstract}

\newpage

\tableofcontents

\newpage

\section{Introduction}

The continuous limit of two-dimensional (2D) critical statistical
systems exhibits conformal invariance which proved to be a very
useful and computationally powerful concept \cite{polyakov,
Belavin:1984vu, cft}. Because of the conformal invariance a system
with a boundary can be studied in any of the topologically
equivalent domains. We will consider critical systems in domains
with the topology of an annulus which for the ease of computations
can be mapped by a conformal transformation onto the upper half
plane with a puncture.

Critical 2D systems can be described in terms of fluctuating curves
(loops, if they are closed) which can be understood as domain walls
or boundaries of clusters (e.g. Fig. \ref{fig:Ising}). The exact
definition depends on the system in question and typical examples
include the boundaries of the Fortuin-Kasteleyn clusters in the
$q$-state Potts model, the loops in the $O(n)$ model, and the
cluster boundaries in a critical percolation system. We use the term
critical curves to describe any of these objects.
\begin{figure}[t]
\centering
\includegraphics[width=0.6\textwidth]{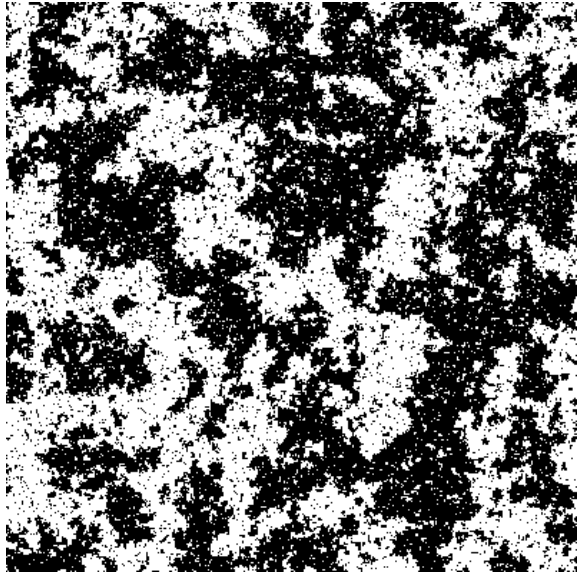}
\caption{A critical Ising model. Black and white represent $\pm 1$
spins. The cluster boundaries furnish an example of critical curves
(after http://www.ibiblio.org/e-notes/Perc/contents.htm).}
\la{fig:Ising}
\end{figure}

Critical curves are drawn from a statistical thermal ensemble and
because of the scale invariance at criticality they are self-similar
random fractals. A large scope of their geometrical properties can
be investigated. As an example, one can study the scaling of the
electric field near a critical curve which is assumed to be the
boundary of a conducting and charged cluster. Averaged over the
fluctuations of the curve (we denote such averages by $\ls \dots
\rs$), the moments of the electric field $E$ scale with the distance
$r$ to the curve as $\ls E^h(r)\rs \sim C(h) r^{\Delta(h)}$ with a
non-trivial multifractal exponent $\Delta(h)$ first computed in
\cite{Duplantier-PRL, duplantier}. On the one hand, this exponent is
related to the multifractal spectrum of the harmonic measure of the
critical curve (and, in particular, determines its fractal
dimension). On the other hand, $\Delta(h)$ happens to be the
gravitationally dressed dimension $h$ introduced in
\cite{Knizhnik:1988ak}, thus forming an interesting link to quantum
gravity. The determination of the $h$-dependent prefactor $C(h)$ is
also interesting and, to our knowledge, has not been done yet. One
can also study more complicated correlators of moments of the
electric field measured at different points: $\ls
E^{h_1}(z_1)E^{h_2}(z_2)\ldots\rs$ (we use complex coordinates $z$
to denote points in the plane). Such objects probe local properties
of critical curves.

In this paper, which extends our results published in Ref.
\cite{BRGW-harmonic-measure-PRL}, we consider both the CFT and SLE
approaches to stochastic geometry of critical curves and elaborate
on the relation between the two. We also obtain the results of
\cite{Duplantier-PRL, duplantier} for multifractal exponents using
CFT without reference to quantum gravity.

\begin{figure}[t]
\centering
\includegraphics[width=0.6\textwidth]{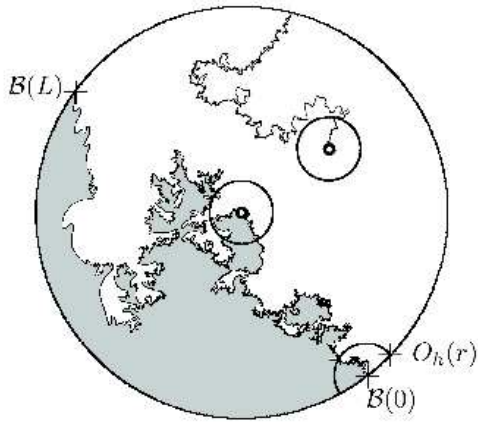}
\caption{A critical curve is conditioned to start from $0$, pass
through a puncture and end at $L$ on the boundary. Another critical
curve is conditioned to end at a puncture in the bulk.}\la{cluster}
\end{figure}

\section{Creation of critical curves and conformal invariance}

As the statistical system fluctuates critical curves come and go in
a random fashion and this makes them difficult to study. We can,
however, choose a point in the bulk or on the boundary and remove
from the partition function all realizations except for those in
which there is a curve passing through this point. The physical
meaning of such conditioning is transparent. Consider a critical
Ising model in the upper half plane and define critical curves to be
the boundaries of clusters of spins of the same sign. As a boundary
condition, we demand that all spins on the negative real axis are
$+1$, and all those on the positive real axis are $-1$. This
condition guarantees that in all realizations there will be a
critical curve growing from the origin.

This is just one of many imaginable ways of conditioning the system
but it illustrates the point: the existence of a curve growing from
a point on the boundary is ensured by a change of the boundary
condition at this point. To condition the curve to pass through a
bulk point one can insert there a puncture, thus effectively turning
it into a boundary point (Fig. \ref{cluster}).

Critical curves are conformally invariant
\cite{polyakov,Belavin:1984vu} (and are sometimes also called
conformally invariant curves) in the sense depicted in Fig.
\ref{fig:conformal invariance}. Define a system in two different but
topologically equivalent domains $A$ and $B$ and introduce a
conformal transformation $f(z)$ which maps $A$ onto $B$. We consider
the ensemble of loops in both domains and also the ensemble $C$ of
images of loops produced by the conformal map $f(z)$. The lattice in
$C$ is warped by the conformal transformation which makes it
different from the system in $B$. Nevertheless, it may happen that
the loops in $B$ and $C$ are statistically identical to each other,
that is, they are different random realizations drawn from the same
statistical ensemble. In this case we say that the ensemble of loops
possesses conformal invariance.
\begin{figure}[t]
\centering
\includegraphics[width=1\textwidth]{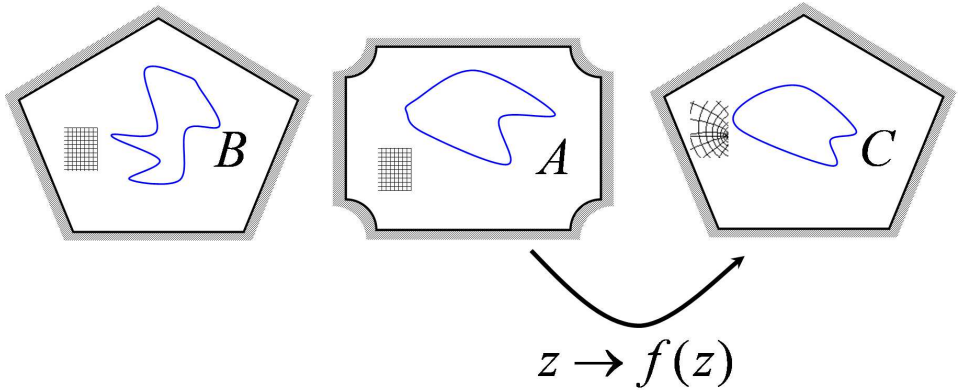}
\caption{$A$ and $B$ are critical systems defined in different
domains. The system $C$ is the image of $A$ under the conformal
transformation $f(z)$ which maps domain $A$ onto domain $B$.
Conformal invariance is the statistical identity of $B$ and $C$.}
\la{fig:conformal invariance}
\end{figure}

The continuous limit of critical systems with conformal invariance
is described by conformal field theory (CFT) which is characterized
by a central charge $c$. In this paper we consider only theories
with $c\leqslant 1$. For $-2\leqslant c\leqslant 1$ they can be
obtained as a continuous limit of the lattice $O(n)$ model.

The central charge itself does not completely determine the critical
curves: their behavior crucially depends on boundary conditions. A
CFT with $c\leqslant 1$ can be represented as a theory of a Bose
field with the Neumann or the Dirichlet boundary conditions. From
the point of view of the $O(n)$ model, different boundary conditions
correspond to dense and dilute phases in which the behavior of
critical curves is different.
\begin{figure}[t]
\centering
\includegraphics[width=0.6\textwidth]{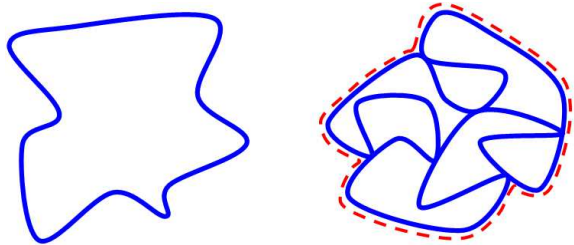}
\caption{Schematic behavior of dilute (left) and dense (right)
loops. The external perimeter of the dense curve is a dilute curve.}
\la{fig:dense dilute}
\end{figure}
An impressive progress in understanding stochastic geometry of
critical curves was achieved in the stochastic Loewner evolution
(also known as Schramm-Loewner evolution, or SLE)
\cite{Schramm}-\cite{Werner2}. It can be deduced that curves in the
theory with the Dirichlet boundary condition (dilute phase) are
almost surely simple. This means that the probability for them to
touch themselves or each other vanishes. Those with the Neumann
boundary condition for $c\geqslant -2$ (dense phase) almost surely
touch themselves and each other (Fig. \ref{fig:dense dilute}). For
$c<-2$ the curves are plane-filling.

There exists a duality relation between dense and dilute phases
\cite{nienhuis}. In particular, the external perimeter of a dense
curve is a dilute curve \cite{duplantier}. Thus, studying the dilute
phase also answers some questions in the dense one. In CFT this
duality is expressed by the fact that both phases can have the same
central charge and the difference lies in the boundary conditions.

\section{Boundary operators and stochastic geometry}
\la{twostepaverage} Consider a system in the dilute phase occupying
the upper half plane and conditioned to contain a critical curve
$\gamma$ connecting the points 0 and $L$ on the real axis (see Fig.
\ref{cluster}). As was mentioned above, this will be the case if the
boundary condition changes at these two points. Such a change is
produced by the insertion of a special boundary condition changing
operator which we can call a curve-creating operator. In statistical
models with $c\leqslant 1$ such as the $O(n)$ model it is known
\cite{cardyoperators} and we will discuss it later. The correlator
of two such operators \ba
\langle\mathcal{B}(0)\mathcal{B}(L)\rangle_{\mathbb{H}} =
\frac{Z(0,L)}{Z}.\nonumber
\end{align}
is determined by the partition function $Z(0,L)$ with boundary
conditions changing at $0$ and $L$. Here $Z$ is the partition
function of the system with uniform boundary conditions. The
subscript of the correlator always refers to the domain of
definition. We note in passing that by fusing together several such
$\mathcal{B}$'s we obtain other boundary condition changing
operators whose insertion produces multiple curves growing from a
point on the boundary.

This correlation function can be interpreted in terms of stochastic
geometry of the curve. The interpretation involves a two-step
averaging (similar arguments can be identified in
\cite{Bauer-Bernard-big-review, recommended}). In the first step we
pick a particular realization of the curve $\gamma$. Then $\gamma$
is the boundary separating two independent systems --- the interior
and the exterior of $\gamma$. In both these systems we can sum over
the microscopic degrees of freedom to obtain the partition functions
$Z^\mathrm{int}_\gamma$ and $Z^\mathrm{ext}_\gamma$, respectively.
These are stochastic objects that depend on the fluctuating geometry
of $\gamma$.

In the second step we average over the ensemble of curves of
$\gamma$. We thus obtain \ba Z(0,L) =\,\,\, \ls
{Z^\mathrm{int}_\gamma Z^\mathrm{ext}_\gamma} \rs,\nonumber
\end{align}
where $\ls\ldots\rs$ stands for averaging over the shape of
$\gamma$.

We further consider the insertion of an additional field
$\mathcal{O}(z)$ on the boundary and a distant field $\Psi(\infty)$
such as to make non-zero the correlation function \ba \la{thecorr}
\langle\mathcal{B}(0)\mathcal{O}(z)\mathcal{B}(L) \Psi(\infty)
\rangle_{\mathbb{H}}
\end{align}
The exact form of $\Psi(\infty)$ is not important, for example we
can choose it as $\mathcal{O}(\infty)$. If we are only interested in
the $z$-dependence of this correlation function in the limit $|z|\ll
|L|$, we can fuse together the distant fields: $\mathcal{B}(L)
\Psi(\infty) :\to \Psi(\infty)$ and therefore consider a three-point
function \ba \langle\mathcal{B}(0)\mathcal{O}(z)\Psi(\infty)
\rangle_\mathbb{H},\nonumber
\end{align}

In the two-step averaging procedure we rewrite (\ref{thecorr}) as an
average over the fluctuating geometry of $\gamma$: $ \la{7} \ls
{\langle \mathcal{O}(z) \Psi(\infty) \rangle_\gamma
Z^\mathrm{int}_\gamma Z^\mathrm{ext}_\gamma} \rs, $ where $\langle
\mathcal{O}(z)\Psi(\infty)\rangle_\gamma$ stands for the correlation
function in the exterior of $\gamma$. In the limit $|z|\ll |L|$ this
correlation function is statistically independent from the other two
factors, and we are left with the scaling relation:
\ba\la{tralalala} \langle\mathcal{B}(0)\mathcal{O}(z)\mathcal{B}(L)
\Psi(\infty) \rangle_{\mathbb{H}}\sim\;\ls \langle
\mathcal{O}(z)\Psi(\infty)\rangle_\gamma\rs\end{align}

The two-step averaging can be extended to multiple operators. If we
choose $\mathcal{O}(z)$ a primary field, this relation yields the
statistics of harmonic measure of critical curves (see Sec.
\ref{sec:harmonic measure}).

\section{Loop models and the Bose field}
\la{sec:CFT}
\subsection{Dense phase}\la{densephase}
The relation between critical curves and operators of a CFT with a
boundary is most transparent in the representation by a Bose field
$\varphi(z,\bar z)$ \cite{cft,nienhuis,Schulze,kawai}. This
representation is commonly known as the Coulomb gas method. As in
the case of any field theory, its derivation from the lattice model
is not rigorous but is {\it a posteriori} justified by the results.

A large variety of 2D lattice statistical models (percolation,
$q$-state Potts, $O(n)$, $XY$, the solid-on-solid and other similar
models) can be defined in terms of a system of random curves,
usually closed loops unless special boundary conditions are imposed.
Each loop has a fixed statistical weight in the ensemble. Mappings
between specific lattice models and loop models are described in
detail in many reviews \cite{nienhuis, loop models 1, loop models 2,
loop models 3}. Here we consider the $O(n)$ model as a
representative example. It covers a large range of central charges
($-2\leqslant c\leqslant1$) and its dilute and dense phases
correspond to the Dirichlet and Neumann boundary conditions of CFT.
The model can be defined on the honeycomb lattice on a cylinder
(such as a carbon nanotube) in terms of closed loops (Fig.
\ref{fig:same topology}): \ba Z^{\textrm{cyl}}_{O(n)} &=
\sum_{\textrm{loops}} x^L n^N. \la{O(n) partition}
\end{align}
Here $x$ is a parameter related to the temperature, $L$ is the total
length of all loops and $N$ is the number of loops. The parameter
$n$ is not necessarily an integer.

The $O(n)$ model has a critical point at
$$x=x_c(n)=\frac{1}{\sqrt{2+\sqrt{2-n}}}$$ for all $n$ in the range $-2 \leqslant
n \leqslant 2$. At the critical point the typical length of a loop
diverges but the loops are dilute in the sense that the fraction of
the vertices visited by them is zero. For $x > x_c(n)$ the typical
length of the loops still diverges so that they are still critical
but visit a finite fraction of the sites. This is the dense phase of
the model. In the continuous limit dilute loops are simple and dense
ones can touch themselves and each other. Finally, at zero
temperature ($x = \infty$) the loops go through every point on the
lattice. This is the fully packed phase. In the continuous limit the
loops become plane filling.

The continuous description of the critical behavior of the $O(n)$
model by a local field theory is best understood in the dense phase.
We need to have a description of the system in terms of local
weights on the lattice and this is done as follows. One randomly
orients the loops and assigns a weight to each orientation. In order
to reproduce the partition function (\ref{O(n) partition}) the sum
of the weights for two orientations of every loop should be $n$.
This is achieved by giving a local complex weight $e^{\pm i e_0
\pi/6}$ to each lattice site where an oriented loop makes a right
(left) turn. The weight of an oriented closed loop is the product of
weights on all sites that it visits. For a closed loop which does
not wrap around the cylinder this product equals $e^{\pm i e_0 \pi}$
since the difference between the numbers of right and left turns is
$\pm 6$. The sum over the two orientations gives the real weight $n$
for an un-oriented loop if we choose \ba n = 2 \cos \pi
e_0.\nonumber
\end{align}
The range of $-2 \leqslant n \leqslant 2$ is covered once by
$0\leqslant e_0\leqslant 1$. To describe both the dilute and the
dense phases, however, we will need to allow for a wider range
$-1\leqslant e_0\leqslant1$, with positive $e_0$ for the dense phase
and negative $e_0$ for the dilute phase.

The situation is different for the loops which wrap around the
cylinder. For them the difference between the numbers of right and
left turns is 0 and they are wrongly counted with weight 2 instead
of $n$. To correct this, one defines for each configuration of
oriented loops a real height variable $H$ which resides on the dual
lattice and takes discrete values conventionally chosen to be
multiples of $\pi$. We start at some reference point where we set
$H=0$, and then every time we cross an oriented loop, we change $H$
by $\pm \pi$ depending on whether we cross the loop from its left to
its right or vice versa. Then one introduces an additional factor
$e^{ie_0(H(+\infty)-H(-\infty))}$ into each term of the partition
function, where $\pm\infty$ stand for two chosen points on the two
boundaries of the system (the bases of the cylinder). This factor is
1 if there are no wrapping loops, so it does not affect the weights
of non-wrapping loops. But since each wrapping loop changes
$H(+\infty)-H(-\infty)$ by $\pm\pi$, this factor turns into the
proper weight $n$ for each wrapping loop. Thus, we represent the
partition function (\ref{O(n) partition}) in the form \ba
Z^{\textrm{cyl}}_{O(n)}= \sum_{\textrm{oriented
loops}}x^Le^{ie_0(H(+\infty)-H(-\infty))}\prod_{\textrm{sites}}e^{\pm
ie_0\pi} . \la{O(n) partition1}\end{align} We note that the
insertion of the fields at the boundaries is brought about not only
by the topology of the domain: in a flat annulus, although its
topology is the same as cylinder's, all loops are counted with the
weight $n$ as it is, so that no extra factors are needed (Fig.
\ref{fig:same topology}).
\begin{figure}[t]
\centering
\includegraphics[width=1\textwidth]{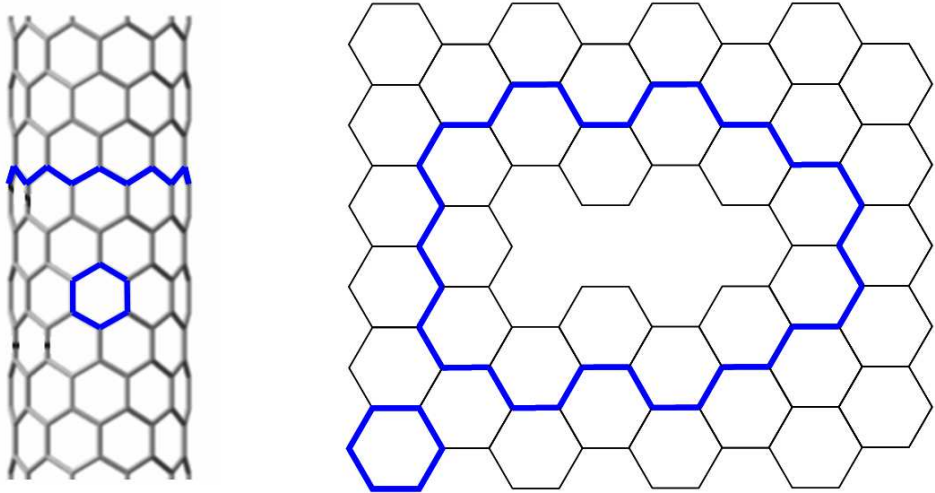}
\caption{If a loop wraps around the cylinder, the difference between
the numbers of left and right turns is 0, otherwise it is $\pm 6$.
On a flat domain of the same topology --- annulus, --- this
difference is $\pm 6$ for all loops.} \la{fig:same topology}
\end{figure}

When we pass to the continuous limit, the lattice height function
$H$ is coarse-grained and becomes a fluctuating compactified scalar
field $h(z,\bar z)$: \ba h\simeq h+2\pi.\nonumber\end{align} Upon
coarse-graining the $O(n)$ loops on a cylinder become level lines of
the field $h$. In the dense phase this field is believed to be
Gaussian, that is, its action contains the term $(g/4\pi)\int d^2x\,
(\nabla h)^2$, where the fluctuation strength parameter $g$ is yet
to be determined. The boundary terms of (\ref{O(n) partition1}) have
to be added in any geometry in the form of an oriented boundary
integral $(ie_0/2\pi)\int_{\der D}dl\, K h$ where $K$ is the
geodesic curvature of the boundary. This is consistent with
(\ref{O(n) partition1}).

A similar situation occurs if the critical system lies on a surface
with curvature which microscopically can be viewed as existence of
defects on the honeycomb lattice (pentagons and heptagons correspond
to positive and negative curvature, correspondingly). The correct
weight for a loop that surrounds a region of non-zero Gaussian
curvature $R$ is obtained if we include in the action the so-called
background charge term $(ie_0/8\pi)\int_Dd^2x\, R h$.

The determination of the coupling constant $g$ can be done either by
comparison with the exact solutions of a six-vertex model related to
the $O(n)$ model or by an elegant argument of \cite{kondev 1, kondev
2}. Another necessary term in the action is the {\it locking
potential} of the form $\lambda \int d^2x\,V(h)$ which favors the
discrete values $h = 0,\pi$. It must be, therefore, a $\pi$-periodic
function of $h$, the most general form of it being $V=\sum_{l \in
{\mathbb Z}, l \neq 0} v_l e^{2ilh}$. We treat the locking potential
perturbatively. In the zeroth order ($\lambda=0$), each term of $V$
is a vertex operator of dimension \cite{nienhuis} $x_l=(2/g)l(l -
e_0).$ An operator with $x_l<2$ is relevant. For the perturbative
approach to be applicable, the locking potential should contain no
relevant terms. Furthermore, the most relevant term should be
marginal. This fixes $g$. For $0 < e_0 < 1$ the most relevant term
is $l=1$ and this gives the relation \ba e_0 &= 1 - g, & n = -2\cos
\pi g. \la{n-g-relation}
\end{align}
Note that $0 < g < 1$, as it is expected in the dense phase.

This description is not valid in the dilute phase. There $H$ does
not renormalize into a Gaussian field. A manifestation of this is
that it is impossible to make the locking potential marginal for $1
\leqslant g \leqslant 2$. Naively, this requires taking $-1 < e_0 <
0$, but then one needs to pick $l = -1$ term as the most relevant
and it still gives us $g = 1 + e_0 = 1 - |e_0| < 1$.

The point $g = 1$ separates the phases and is somewhat special:
there the $l = 1$ and $l= -1$ terms in the locking potential have
the same dimension and we have to keep them both. The relation
between Gaussian fields and critical curves has been made rigorous
in \cite{SS-gaussian-field} which in our language describes the case
$n = 2$, $g = 1$.

We now introduce the parametrization \ba g =
\frac{4}{\kappa}.\nonumber
\end{align}
The range $0 < g \leqslant 2$ where the $O(n)$ model exhibits
critical behavior corresponds to $2 \leqslant \kappa < \infty$.
Smaller values $0 < \kappa < 2$ correspond to multi-critical points
in the $O(n)$ model but also make sense for other critical systems.
Notice that $\kappa < 4$ and $\kappa > 4$ describe the dilute and
the dense phases correspondingly, while $\kappa = 4$ gives the point
$g = 1$ separating the two phases. Readers familiar with SLE will
recognize that this is similar to the SLE phases
\cite{Rohde-Schramm}. In fact, $\kappa$ can be identified with the
SLE parameter. For $\kappa\leqslant 4$ SLE curves are simple and for
$4<\kappa<8$ they have double points.

In the literature it is customary to rescale the field $\varphi =
\sqrt{2g} h$ and fix the coupling constant $g$ to be $1/2$ at the
expense of varying the compactification radius of $\varphi$: \ba
{\mathcal R} = \sqrt\frac{8}{\kappa}. \la{comp radius}
\end{align}
This is the normalization that we adopt from now on. Up to the
locking potential one then writes the partition function as
\ba\la{cosinus}
Z_{O(n)}\sim\int\mathcal{D}\varphi\cos\Bigl[\frac{\sqrt2\alpha_0}{2\pi}\oint_{\der
D}dl\,
K\varphi\Bigr]e^{-\int(\nabla\varphi)^2\frac{d^2x}{8\pi}},\end{align}
where the relation (\ref{n-g-relation}) reads\footnote{The notation
differs from \cite{BRGW-harmonic-measure-PRL}, where $\alpha_0<0$ in
the dense phase. As a consequence, in the Kac notation $\psi_{r,s}$
of curve-creating operators $r\leftrightarrow s$.} \ba 2\alpha_0 =
\frac{\sqrt\kappa}{2} - \frac{2}{\sqrt\kappa}>0.\nonumber\end{align}
The marginal locking potential now reads $\lambda \int
d^2x\,e^{i\sqrt2\alpha_+\varphi}$, where we used the notation \ba
\alpha_+ = \frac{\sqrt\kappa}{2}, \quad \alpha_- = -
\frac{2}{\sqrt\kappa}. \la{alphas}
\end{align}
The partition function written in this way remains valid for any
flat geometry with any number of boundaries.

Note that $\alpha_\pm$ satisfy simple relations $\alpha_\pm =
\alpha_0 \pm \sqrt{\alpha_0^2 + 1}$. The locking potential is
treated as a perturbation: in each order of the perturbation theory
a number of screening charges $\int
d^2x\,e^{i\sqrt2\alpha_+\varphi}$ is inserted into all correlation
functions so as to satisfy the neutrality condition \cite{df}. Note
that the second screening charge $\int d^2x\,
e^{i\sqrt2\alpha_-\varphi}$, although also of zero dimension, is not
found by the argument on the lattice --- it is incompatible with the
compactification radius.

If the domain's Gaussian curvature $R$ is not zero (as it happens
when heptagons or pentagons are inserted as defects of the lattice)
an additional term $i\frac{\sqrt2\alpha_0}{4\pi}\int d^2x\,R\varphi$
is needed.

Due to the symmetry $\varphi\to -\varphi$ we can rewrite the
partition function as a local field theory
$Z_{O(n)}\sim\int\mathcal{D}\varphi e^{-S}$, but with a {\it
complex} action $S=\mathcal{A}+ \lambda\int_D d^2 x \, e^{i\sqrt{2}
\alpha_ + \varphi}$ where \ba \mathcal{A} = \frac{1}{8\pi} \int_D
d^2 x \,(\nabla\varphi)^2 + i \frac{\sqrt2\alpha_0}{2\pi} \int_{\der
D} dl \, K \varphi,\quad \textrm{dense phase, }\alpha_0>0.
\la{bigaction}
\end{align}
Since this action is local we can use the usual methods of conformal
field theory. Such an action is known to describe a field theory
with central charge \ba c= 1 - 24 \alpha_0^2 = 1 - 3 \frac{(\kappa -
4)^2}{2 \kappa}. \nonumber
\end{align}
However, the saddle point of the action does not lie in the real
fields and therefore the Bose field in the OPEs is not real so that
it is no longer the coarse-grained height function.

\subsubsection{Conformal transformations and boundary conditions}

As an illustration, we consider the upper half plane (with
punctures). In this case the curvature of the boundary is zero
except at the point at infinity. By conformal invariance, the theory
can be mapped onto any domain with a curved smooth boundary.

Since the field is Gaussian it consists of the holomorphic and the
antiholomorphic parts: \ba \varphi(z,\bar z) = \phi(z) +
\bar\phi(\bar z),\nonumber\end{align} whose correlation functions
are \ba \langle \phi(z)\phi(z') \rangle &\!=\! -\!\log(z\! -\!
z'),\;\; \langle \bar\phi(\bar z) \bar\phi(\bar z') \rangle\!=\!
-\!\log(\bar z\! -\! \bar z'),\;\; \langle \phi(z) \bar\phi(\bar z)
\rangle\! =\! 0. \la{propagators}
\end{align}
We will also use the dual boson field $\w\varphi$ related to
$\varphi$ through Cauchy-Riemann conditions: \ba\la{dualfield}
\w\varphi(z,\bar z) = -i\phi(z) + i\bar\phi(\bar z).
\end{align}
Varying the action $\mathcal{A}$ with respect to the metric we find
the components of the stress-energy tensor: \ba T(z) &=
-\frac{1}{2}(\der\phi)^2 + i\sqrt 2 \alpha_0 \der^2 \phi, & \bar
T(\bar z) &= -\frac{1}{2}(\bar\der\, \bar\phi)^2 + i\sqrt 2 \alpha_0
\bar\der^2 \bar\phi, \la{stress-chiral-dense}
\end{align}
where $\der = \der/\der z, \bar\der = \der/\der \bar z$, and the
products are normal ordered.

When the critical system is defined in the upper half plane the
conformally invariant boundary condition is $T(x) = \bar T(x)$ for
all $x \in \mathbb R$ \cite{cardyoperators}. For the stress-energy
tensor (\ref{stress-chiral-dense}) this condition readily yields
$\phi(x) = \bar\phi(x)$, therefore $\varphi=\phi(z)+\phi(\bar z)$
and $\w\varphi=-i\phi(z)+i\phi(\bar z)$ which is equivalent to the
Neumann boundary condition: \ba \der_y \varphi\big|_{y=0} =
0.\nonumber
\end{align}
The Neumann condition implies that $\varphi$ is a scalar which
agrees with the meaning of the Bose field as height function.

The components (\ref{stress-chiral-dense}) determine the
transformation properties of the fields under a conformal
transformation $z \to f(z)$: \ba \phi(z) &\to \phi(f(z)) + i
\sqrt{2} \alpha_0 \log f'(z), \la{phi-transformation-dense}
\end{align}
which yields \ba\varphi(z,\bar z)\to\varphi(f(z),\overline{f(z)})
+i2\sqrt2\alpha_0\log |f'(z)|.\nonumber\end{align} This
transformation is merely the transformation of the classical
configuration of $\varphi$ (the saddle point of (\ref{cosinus})).
Notice that although $\varphi$ is not real, its real part does not
transform so that the level lines \ba \mathrm{Re}\,\varphi(z,\bar
z)=\mathrm{const.}\nonumber\end{align} are conformally invariant
fluctuating curves. This suggests to identify them with the
continuous limit of critical curves i.e. the level lines of the
coarse-grained height function.

\subsubsection{Electric and magnetic operators}
In the following we will see that critical curves in both dense and
dilute phase are created by spinless vertex operators: \ba\la{the
big fat operator} \mathcal{O}^{(e,m)}(z,\bar z) &= e^{i\sqrt2 e
\varphi(z,\bar z)}e^{-\sqrt2 m \w\varphi(z,\bar z)},
\end{align}
where we introduced ``electric" and ``magnetic" charges $e$ and $m$.
The boundary term in the action (\ref{bigaction}) is the insertion
of an electric background charge $e=2\alpha_0$ at a puncture in the
upper half plane.

A vertex operator is a primary field and can be written as a product
of the holomorphic and the antiholomorphic components
$\mathcal{O}^{(e,m)}(z,\bar z)=V^\alpha(z)V^{\bar\alpha}(\bar z),$
which are assumed to be normal ordered: \ba V^\alpha(z) & =
e^{i\sqrt 2 \alpha \phi(z)}, \qquad {\bar V}^{\bar\alpha}(\bar z) =
e^{i \sqrt2 \bar\alpha \bar\phi(\bar z)}. \nonumber
\end{align}
Here the holomorphic and antiholomorphic charges are
\ba\la{chargescharges} \alpha &= e + m, & \bar\alpha = e - m.
\end{align}
In the dense phase the Neumann boundary condition allows to write
\ba \mathcal{O}^{(e,m)}(z,\bar z)&= e^{i \sqrt{2} (e + m) \phi(z)}
e^{i \sqrt{2} (e - m) \phi(\bar z)}. \la{parity}
\end{align}
This can be interpreted as gluing together the holomorphic and the
antiholomorphic sectors which is best understood in terms of image
charges \cite{cardyoperators, Schulze, kawai}. The operator
(\ref{parity}) is regarded as a product of two holomorphic
operators: one in the upper half plane and the other at the image
point in the lower half plane. In the image the electric charge
remains the same while the magnetic charge changes sign.

The holomorphic and the antiholomorphic weights of the vertex
operators are found by applying the stress-energy tensor
(\ref{stress-chiral-dense}) to them: \ba h_\alpha &= \alpha(\alpha -
2\alpha_0), & {\bar h}_{\bar \alpha} &= \bar\alpha (\bar\alpha -
2\alpha_0).\nonumber
\end{align}
A given weight $h$ corresponds to two charges: $
\alpha_0\pm\sqrt{\alpha_0^2+h}$.

In terms of the electric and magnetic charges the weights read \ba
h(e,m) = (e + m)(e + m - 2\alpha_0), && {\bar h}(e,m) = (e - m)(e -
m - 2\alpha_0). \la{e-m-dimensions-dense}
\end{align}
A vertex operator is spinless (meaning that $h = \bar h$) if either
$\bar\alpha = \alpha$ or $\bar\alpha = 2\alpha_0 - \alpha$. In the
first case the operator is purely electric ($m=0$) and in the second
case it can have an arbitrary magnetic charge $m$ but the electric
charge should be $e = \alpha_0$.

We will also find it convenient to denote the vertex operators by
their weights. Thus $\mathcal{O}_h(z,\bar z)$ is a spinless vertex
operator of weight $h$ in the bulk and $\mathcal{O}_h(x)$ is a
boundary vertex operator of weight $h$.

The propagators (\ref{propagators}) lead to fusion by the addition
of charges: \ba\la{fusionrule} \lim_{z_1\to
z_2}V^{\alpha_1}(z_1)V^{\alpha_2}(z_2) =
(z_1-z_2)^{2\alpha_1\alpha_2}V^{\alpha_1+\alpha_2}(z_1)+\ldots.\end{align}

In the literature it is customary to label holomorphic charges and
weights by two numbers $r,s$ according to \ba \alpha_{r,s} &=
\alpha_0 - \frac{1}{2}(r \alpha_+ + s \alpha_-) =
\frac{1}{2}(1-r)\alpha_+ + \frac{1}{2}(1-s)\alpha_-, \nonumber \\
h_{r,s} &= \alpha_{r,s}(\alpha_{r,s} - 2\alpha_0) = \frac{(r \kappa
- 4 s)^2 - (\kappa - 4)^2}{16 \kappa}. \la{Kac weights}
\end{align}
The primary field of weight $h_{r,s}$ is denoted $\psi_{r,s}$. We
will use it as a shorthand notation without imposing any
restrictions on $r$ or $s$.

\subsubsection{Curve-creating operators in the dense phase}

In this section we determine the vertex operators which create
critical curves.

On the lattice, the appearance of $n$ critical curves emanating from
a small region in the bulk results in the change in the height
function $H$ by $\pm \pi n$ as one makes a full circle around this
region. This assumes that all of the curves are oriented the same
way --- inwardly or outwardly, --- for otherwise they would be able
to reconnect with each other. In the continuous limit
$H\to\mathrm{Re}\,\varphi$ this represents a vortex configuration of
the field $\varphi$. As the propagators (\ref{propagators}) show, a
vortex is produced by the insertion of a magnetic charge: \ba
\mathcal{O}^{(e,0)}(z, \bar z) \mathcal{O}^{(0,m)}(z', \bar z') &=
\Big( \frac{z - z'}{\bar z - \bar z'} \Big)^{2em}
\mathcal{O}^{(e+m)}(z') \mathcal{O}^{(e-m)}(\bar
z') + \ldots \nonumber \\
&= e^{4i e m \arg(z - z')} \mathcal{O}^{(e+m)}(z')
\mathcal{O}^{(e-m)}(\bar z') + \ldots \nonumber\end{align} This
means that for $z\to z'$ the value of the field changes by $\pm
4\sqrt2\pi m$ as one goes around $z'$ in full circle. Hence, a
discontinuity line arises (Fig. \ref{fig:bulkmagnet}).
\begin{figure}[t] \centering
\includegraphics[width=0.6\textwidth]{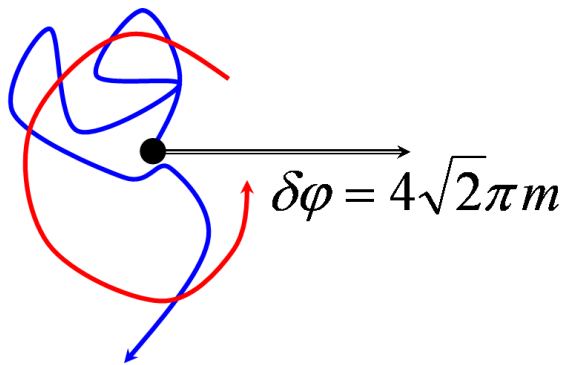}
\caption{A bulk magnetic operator $\mathcal{O}^{(0,m)}$ creates a
vortex configuration of the field $\varphi$. One critical curve is
shown. The double line is the discontinuity line of the field.}
\la{fig:bulkmagnet}
\end{figure}
If this vortex corresponds to a star of $n$ critical curves joining
at the point $z'$, the change in $\varphi$ should be equal to $\pm n
\pi {\mathcal R}$. The compactification of $\varphi$ makes the
discontinuity line invisible for an even number of curves. For an
odd number of curves, a discontinuity line with the jump $\pi
{\mathcal R}$ remains. One can view it as a critical curve whose
form is conditioned to be fixed so that it is removed from the
ensemble of random curves.

We then find the magnetic charge of the bulk operator which creates
$n$ curves as \ba m = \pm \frac{\sqrt{2}}{8} n {\mathcal R} = \pm
\frac{n}{2 \sqrt{\kappa}} = \mp \frac{n}{4} \alpha_-, \nonumber
\end{align}
where we used the value (\ref{comp radius}) of the compactification
radius and the definition (\ref{alphas}) of $\alpha_-$. In order to
be spinless (otherwise the operator would transform under rotations,
giving non-trivial dependence on the winding number of curves), the
bulk curve-creating operator should also have an electric charge
$\alpha_0$ thus acquiring the form \ba V^{\alpha_0 + m}(z)
V^{\alpha_0 - m}(\bar z).\nonumber\end{align} We have therefore
found the possible holomorphic charges of the bulk curve-creating
operator to be \ba \alpha &= \alpha_0 \mp \frac{n}{4} \alpha_- =
\alpha_{0,\pm n/2}\nonumber
\end{align}
in the notation (\ref{Kac weights}). These charges correspond to the
same weight \ba h_{0,n/2} = \frac{4n^2 - (\kappa - 4)^2}{16 \kappa}.
\nonumber
\end{align}
The curve-creating operator itself is then denoted $\psi_{0,n/2}$.
In particular, a single critical curve going through a point $z$
(which is the same as $n = 2$ critical curves meeting at the point
$z$) is created by the operator $\psi_{0,1}$ with the holomorphic
weight $h_{0,1} = (8 - \kappa)/16$. Note that this weight determines
the fractal dimension of a critical curve as $2 - 2h_{0,1}$
\cite{BB-zigzag}.

The creation of curves on the boundary is considered in the same
way. If $n$ curves exit a boundary point the value of the height
function changes by $\pm n\pi\mathcal{R}$ as one goes around this
point in a semi-circle. On the real axis $[e^{\pm
i\sqrt2\alpha\phi(x)},\phi(y)]=\pm\sqrt2\pi\alpha\,\theta(x-y)e^{\pm
i\sqrt2\alpha\phi(x)}$ where $\theta(x)$ is the step-function. This
shows that the field value $\phi$ on the boundary changes by $\pm
\sqrt2\pi\alpha$ in crossing the inserted vertex operator. Moreover,
in presence of the Neumann boundary condition $\varphi(x)=2\phi(x)$
on the real axis. The $n$ curves are created, therefore, by the
insertion of a boundary operator of the form $e^{\pm
i\sqrt2\alpha_{1,n+1}\phi}$. The weight depends on the sign in the
exponent. The lower weight corresponding to plus sign so that this
operator is more relevant. This is then the curve-creating operator
on the boundary denoted $\psi_{1,n+1}$. There also exist operators
which condition a curve not to touch a certain part of the boundary
\cite{BB-zigzag, bauer-bernard-1, bauer-bernard-2, bauer-bernard-5}.

\subsection{Dilute phase $\alpha_0<0$}
\la{dilute phase}

Dilute critical curves occur, for example, in the $O(n)$ model at
the critical temperature. They are known to be described by $g>1$
($\kappa<4$). The central charge and the conformal weights of all
curve-creating operators are obtained by continuation of the
corresponding formulas from the dense phase ($\kappa>4$). The
arguments used to derive the action of field theory from the lattice
in the dense phase cannot be straightforwardly extended to the
dilute phase. Nevertheless, the description of the dilute phase is
available via the electric-magnetic duality transformation
$e\leftrightarrow m$ also called $T$-duality in the literature (see
\cite{duplantier, nienhuis, kadanoff, polchinski}). Every
correlation function of primary fields
--- and in particular curve-creating operators, --- can be
represented as the insertion of electric and magnetic charges into
the partition function: $Z[g,e,m]$. The electric-magnetic duality
means $Z[g,e,m]=Z[1/g,m,e]$.

The interchange between the electric and magnetic background charge
means the following change of the boundary term in the action
(\ref{bigaction}):\ba i \frac{\sqrt2\alpha_0}{2\pi} \int_{\der D} dl
\, K \varphi\to -\frac{\sqrt2\alpha_0}{2\pi}\int_{\der D}dl\,
K\w\varphi,\nonumber\end{align} where $\w\varphi$ is the Hilbert
transform of $\varphi$ on the boundary of the domain $D$, with the
property $\der_l\w\varphi\big|_{\der D}=-\der_n\varphi\big|_{\der
D}$. The normal derivative is taken in the direction inside $D$.
Once $\varphi$ is harmonic $\w\varphi$ coincides with the definition
(\ref{dualfield}). The electric-magnetic transformation makes
$\varphi$ a pseudo-scalar. As a result the action reads
\ba\la{diluteaction} \mathcal{A}=\frac{1}{8\pi} \int_D d^2 x
\,(\nabla\varphi)^2 -\frac{\sqrt2\alpha_0}{2\pi}\int_{\der D}dl\,
K\w\varphi, \quad\textrm{dilute phase, }\alpha_0<0.
\end{align}

We note that in the dilute phase the action is real and its saddle
point lies in the real fields. The connection of the field $\varphi$
to the critical curves in the dilute phase will be discussed below.

For simplicity, we start with the theory in the upper half plane.
The components of the stress-energy tensor now read (cf.
(\ref{stress-chiral-dense})) \ba T(z) &= -\frac{1}{2}(\der\phi)^2 +
i\sqrt 2 \alpha_0 \der^2 \phi, & \bar T(\bar z) &= -\frac{1}{2}
(\bar\der\, \bar\phi)^2 - i\sqrt 2 \alpha_0 \bar\der^2 \bar\phi.
\la{stress-chiral-dilute}
\end{align}
The condition $T(x)=\bar T(x)$ now glues the holomorphic and
antiholomorphic sectors on the real axis as $\phi(x)=-\bar\phi(x)$
which means $\varphi(z,\bar z)=\phi(z)-\phi(\bar z),$ $\w\varphi =
-i\phi(z)-i\phi(\bar z)$ and therefore the boundary condition on the
real axis is Dirichlet:\ba\la{dirichlet}
\der_x\varphi\big|_{y=0}=0.\end{align} Unlike in the dense phase,
$\bar\phi$ is complex conjugate of $\phi$.

The holomorphic and antiholomorphic weights of $\mathcal{O}^{(e,m)}$
in the dilute phase are (cf. (\ref{e-m-dimensions-dense})) \ba
h(e,m) = (e + m)(e + m - 2\alpha_0), && {\bar h}(e,m) = (e - m)(e -
m + 2\alpha_0)\nonumber
\end{align} or \ba h=\alpha(\alpha-2\alpha_0),\quad
\bar h = \bar\alpha(\bar\alpha+2\alpha_0)\nonumber\end{align} in the
notation (\ref{chargescharges}). Since in the dilute phase
$\alpha_0<0$, the charge which corresponds to a weight $h$ and
vanishes with $h$ is \ba\la{alphahdilute}
\alpha_h=\alpha_0+\sqrt{\alpha_0^2+h},\quad \kappa\leqslant
4\end{align} A spinless vertex operator in the dilute phase is
either purely magnetic $e=0$, or it has a magnetic charge
$m=\alpha_0$ and an arbitrary electric charge.

The stress-energy tensor (\ref{stress-chiral-dilute}) shows that
under a conformal transformation $z\to f(z)$ \ba\varphi(z,\bar
z)\to\varphi(f(z),\overline{f(z)}) -2\sqrt2\alpha_0\arg
f'(z),\nonumber\end{align} which preserves the reality of $\varphi$.
We observe that the flow lines of the vector field \ba\la{flowlines}
e^{\frac{i}{2\sqrt2\alpha_0}\varphi}\end{align} are conformally
invariant since (\ref{flowlines}) has conformal spin $h-\bar h=-1$.
If $\varphi$ were a smooth function the flow lines would repulse.
This suggests identifying them with the critical curves in the
dilute phase. The identification of flow lines (\ref{flowlines}) for
the Gaussian field with Dirichlet boundary conditions with
conformally invariant curves in SLE was done in \cite{AC-geometry}.

In the dilute phase the boundary of the domain is a flow line. The
boundary condition (\ref{dirichlet}) transforms with the
transformation of the domain. If $f_D(z)$ maps the domain $D$ onto
the upper half plane (a domain with a straight boundary) the
boundary condition in $D$ is (Fig. \ref{fig:boundary condition})
\ba\la{dilute bc} \varphi\big|_{\der D} &= -2\sqrt2\alpha_0\arg
f'_D\big|_{\der D}.
\end{align}
At one point on the boundary $\varphi$ undergoes a jump $4\sqrt
2\alpha_0\pi,$ which means that a magnetic charge $2\alpha_0$ is
inserted there. In the upper half plane this magnetic charge is
pushed out to infinity.
\begin{figure}[t]
\centering
\includegraphics[width=0.75\textwidth]{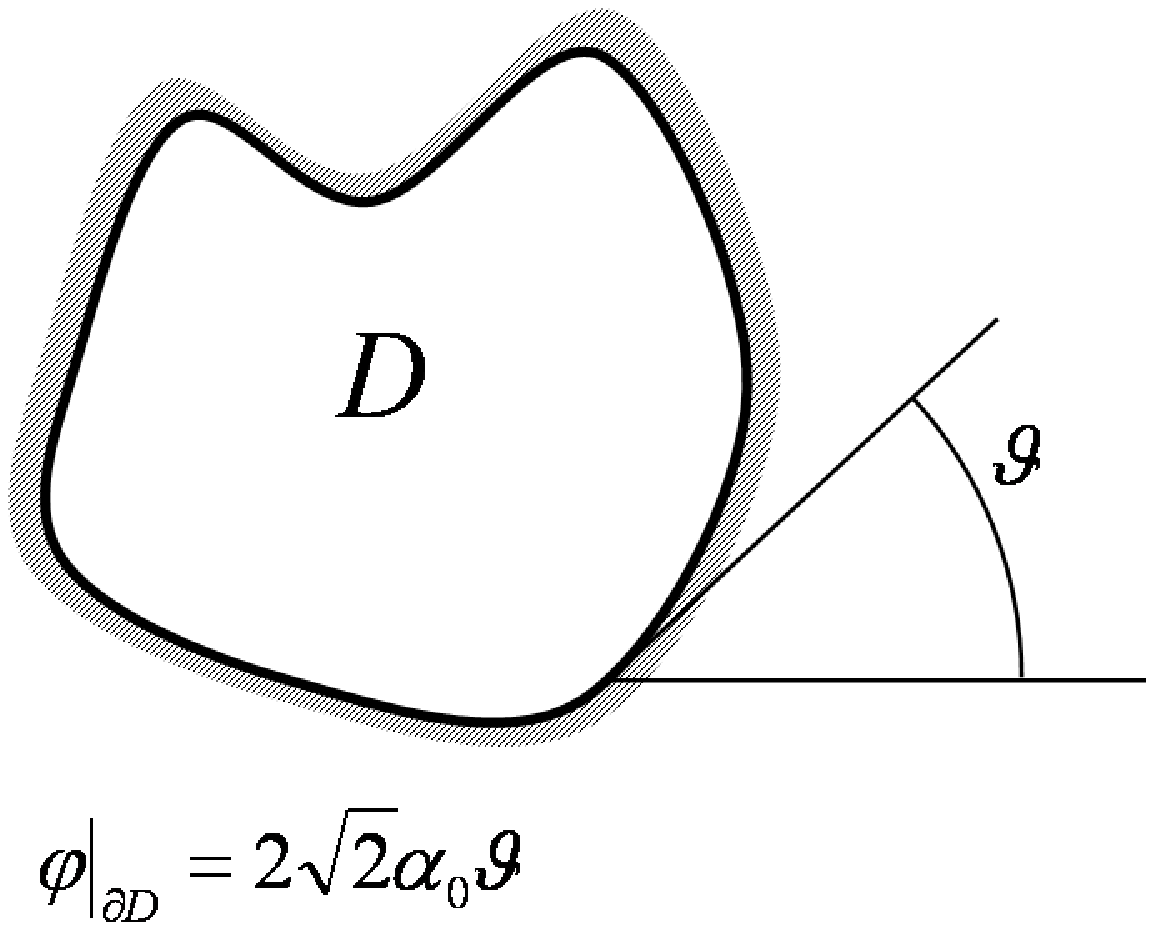}
\caption{Boundary condition (\protect\ref{dilute bc}).
$\vartheta=-\arg f_D'\big|_{\der D}$.} \la{fig:boundary condition}
\end{figure}

In the upper half plane the Dirichlet boundary condition allows us
to write the vertex operator (\ref{the big fat operator}) as (cf.
eq. (\ref{parity})) \ba \mathcal{O}^{(e,m)}(z,\bar z)&= e^{i
\sqrt{2} (e + m) \phi(z)} e^{i \sqrt{2} (m-e) \phi(\bar
z)},\nonumber\end{align} which shows that the image of an electric
charge changes sign and the image of the magnetic charge does not.

The conformal weights and holomorphic charges of the curve-creating
operators are found by the continuation of the corresponding
formulas from the dense phase. That is, the bulk operators giving
rise to $n$ critical dilute curves are $\psi_{0,n/2}$ and
$\psi_{1,n+1}$, but with $\kappa<4$ in the definitions (\ref{Kac
weights}) of their weights. We summarize the operators which create
$n$ curves in the bulk and
on the boundary in both phases in the following table.\\
\\
\begin{tabular}{c||c|c|c||c|c|c|}
\cline{2-7}
 &\multicolumn{3}{c||}{$n$ curves in bulk} &\multicolumn{3}{c|}{$n$ curves on boundary}\\
 \cline{2-7}
&$r,s$ & $e,m$ & explicit & $r,s$ & $\alpha$ & explicit \\
\hline\hline
&  &  &  &  &  &  \\
 Dense & $0,\frac{n}{2}$&
$\alpha_0,\mp\!\frac{n}{4}\alpha_-$ &
$e^{i\sqrt2\alpha_0\varphi}e^{\pm\!\frac{\sqrt2n}{4}\alpha_-\w\varphi}$
& $1,n\!+\!1$ & $\!-\!\frac{n}{2}\alpha_-$ &
$e^{-\!i\frac{n}{\sqrt2}\alpha_-\phi}$\\
&  &  &  &  &  &  \\
\hline
&  &  &  &  &  &  \\
Dilute & $0,\frac{n}{2}$& $\mp\!\frac{n}{4}\alpha_-,\alpha_0$ &
$e^{\mp\!
i\frac{\sqrt{2}n}{4}\alpha_-\varphi}e^{-\!\sqrt2\alpha_0\w\varphi}$
& $1,n\!+\!1$ & $\!-\!\frac{n}{2}\alpha_-$ &
$e^{-\!i\frac{n}{\sqrt2}\alpha_-\phi}$\\
&  &  &  &  &  &  \\
\hline
\end{tabular}
\subsection{Differential equations for the boundary curve-creating operator}

Due to the conformal invariance any correlation function which
includes a boundary curve-creating operator satisfies a differential
equation which we derive below. The correlation functions then can
be found as solutions of this equation. This applies to both the
dense and the dilute phases.

The stress-energy tensor (\ref{stress-chiral-dense}) or
(\ref{stress-chiral-dilute}) is a field which generates
transformations of primary fields under changes of geometry. The
change of the upper half plane into that with a puncture at $w_0$ is
produced by an infinitesimal transformation conformal everywhere
except $w_0$: \ba z \to f(z) &= z + \epsilon \Bigl(\frac{1}{z - w_0}
+ \frac{1}{z- \bar w_0}\Bigr), & \epsilon\to 0.\nonumber
\end{align}
This transformation affects correlation functions of primary fields
in the usual way via the primary transformation law. A correlation
function in the new geometry is \ba &\Big\langle \prod_{i=1}^k
V^{\alpha_i}(z_i) \Big\rangle_{\mathbb{H}\backslash w_0} =
\prod_{i=1}^k \Big( \frac{df^{-1}(z_i)}{d z_i} \Big)^{h_i}
\Big\langle \prod_{i=1}^k
V^{\alpha_i}(f^{-1}(z_i)) \Big\rangle_{\mathbb H} \nonumber \\
& \quad = \Bigl[1+\epsilon\sum_{i=1}^k\Bigl(\frac{h_i}{(z_i-w_0)^2}
- \frac{1}{z_i-w_0}\der_{z_i}+ \{ w_0\to \bar w_0\}\Bigr)\Bigr]
\Big\langle \prod_{i=1}^k V^{\alpha_i}(z_i) \Big\rangle_{\mathbb
H}.\nonumber
\end{align}
The same change of the domain is by definition produced by the
insertion of the stress-energy tensor $T(w_0)+\bar T(\bar w_0)$ into
the correlation function (the conformal Ward identity).

We now fix $w_0$ to be a real number $x$. On the real axis
$T(x)=\bar T(x)$ and the Ward identity reads \ba \langle T(x)
\mathcal{O}(z_1,\ldots z_k)\rangle_{\mathbb{H}} = \mathcal{L}_{-2}
\langle \mathcal{O}(z_1,\ldots z_k)\rangle_{\mathbb{H}},\nonumber
\end{align}
where $\mathcal{O}(z_1,\ldots z_k)=V^{\alpha_1}(z_1) \ldots
V^{\alpha_k}(z_k)$ and \ba \mathcal{L}_{-n} &=
\sum_{i=1}^k\Bigl(\frac{(n-1)h_i}{(z_i-x)^n} -
\frac{1}{(z_i-x)^{n-1}}\der_{z_i}\Bigr). \la{virasororep}
\end{align}
We remark that due to translational invariance the action of $\der_x
+ \sum_{i=1}^k\der_{z_i}$ on the correlation function is zero so
that ${\mathcal L}_{-1} = \der_x $.

The product of fields inside a correlation function is not
normal-ordered. We can also insert the stress-energy tensor at the
same point as one of the fields using normal ordering: \ba \langle
\nol T(x)V^{\alpha}(x) \nor O(z_1,\ldots z_k)\rangle_{\mathbb{H}} =
\mathcal{L}_{-2}\langle V^{\alpha}(x)O(z_1,\ldots
z_k)\rangle_{\mathbb{H}} \la{theverysame}
\end{align}
with the same definition of $\mathcal{L}_{-2}$. For special values
of $\alpha$ this leads to a differential equation
\cite{Belavin:1984vu, cft}.

Using the series expansion $V^\alpha(x) =
\sum_{n=0}^\infty\frac{(i\sqrt2\alpha)^n}{n!} \nol \phi^n(x) \nor$
and the Wick theorem we find the normal-ordered product \ba \nol
\der\phi(x) \der\phi(x) V^\alpha(x) \nor &= \nol \der\phi(x)
\der\phi(x) e^{i\sqrt2\alpha\phi(x)} \nor - i 2\sqrt2 \alpha \nol
\der^2\phi(x) e^{i\sqrt2\alpha\phi(x)} \nor,\nonumber\\ \nonumber
\nol \der^2\phi(x)V^\alpha(x) \nor &= \nol
\der^2\phi(x)e^{i\sqrt2\alpha\phi(x)} \nor.
\end{align}
The second term in the first line comes from only one of
$\der\phi(x)$ being contracted with the exponential. We then find
\ba \nol T(x)V^\alpha(x) \nor = \nol
(-\frac{1}{2}\der\phi(x)\der\phi(x) + i\sqrt2(\alpha_0 +
\alpha)\der^2\phi(x)) e^{i\sqrt2\alpha\phi(x)} \nor .\nonumber
\end{align}
On the other hand, a simple differentiation gives \ba \der^2
V^\alpha(x) = \nol (-2\alpha^2\der\phi(x)\der\phi(x)+
i\sqrt2\alpha\der^2\phi(x))e^{i\sqrt2\alpha\phi(x)} \nor .\nonumber
\end{align}
These two expressions are the same up to a prefactor provided
$4\alpha(\alpha + \alpha_0) = 1$. The two solutions of this equation
are $\alpha_{1,2} = 1/\sqrt\kappa$ and $\alpha_{2,1} =
-\sqrt{\kappa}/4$. They determine operators degenerate on level 2:
$\psi_{1,2}\equiv e^{i\sqrt2\alpha_{1,2}\phi}$ and $\psi_{2,1}\equiv
e^{i\sqrt2\alpha_{2,1}\phi}$. We then write \ba \nol
\Bigl(\frac{\kappa}{4}\der^2-T(x)\Bigr)\psi_{1,2}(x) \nor &= 0, &
\nol \Bigl(\frac{4}{\kappa} \der^2 - T(x)\Bigr)\psi_{2,1}(x) \nor &=
0.\nonumber
\end{align}
These are operator equations, meaning that they can be inserted into
any correlation function leading, for example, to \ba \langle \nol
\Bigl(\frac{\kappa}{4}\der^2_{x} - T(x)\Bigr) \psi_{1,2}(x) \nor
O(z_1,\ldots z_k)\rangle_{\mathbb{H}}=0.\nonumber
\end{align}
If all fields are primary we combine this result with the Ward
identity (\ref{theverysame}) and obtain the following differential
equations: \ba (\frac{\kappa}{4} {\mathcal L}^2_{-1} - {\mathcal
L}_{-2}) \langle \psi_{1,2}(x) O(z_1,\ldots z_k)
\rangle_{\mathbb H} &= 0, \la{cfteq1} \\
(\frac{4}{\kappa} {\mathcal L}^2_{-1} - {\mathcal L}_{-2}) \langle
\psi_{2,1}(x) O(z_1,\ldots z_k)\rangle_{\mathbb{H}} &= 0.\nonumber
\end{align}
The first of these equations is relevant for calculations of
correlation functions containing the boundary curve-creating
operator $\psi_{1,2}$.

\section{Stochastic (Schramm-) Loewner evolution (SLE)}
\la{sec:SLE} In the previous sections correlation functions of
boundary operators were interpreted as averaging over the shapes of
critical curves. The shape of a curve fluctuates with some
probability measure i.e. there is a certain statistical weight
associated with each possible shape. The SLE approach \cite{Schramm}
determines this measure. If a curve is conditioned to a certain
shape objects of field theory explicitly depend on it.

In this section we briefly review SLE
(\cite{Schramm}--\cite{Werner2}) and present the Bose field, its
current, the vertex operators and the stress-energy tensor
explicitly in terms of the shape of the curve. We restrict the
discussion to the dilute case $\kappa\leqslant 4$.
\subsection{Loewner equation}

Consider the upper half-plane $\mathbb{H}$ in which an arbitrary
non-self-intersecting curve $\gamma$ starting from the origin is
drawn. Let the curve be parameterized by $t \geqslant 0$ called
``time''. This setting defines a continuous family of slit domains
each member of the family being the upper half plane with a part of
$\gamma$ up to some value $t$ removed: $\mathbb{H}_t = \mathbb{H}
\backslash \gamma_t.$ Each slit domain $\mathbb{H}_t$ is
topologically equivalent to the upper half plane since $\gamma$ does
not have self-intersections. We define the time-dependent conformal
map $w_t(z)$ of the slit domain to the upper half plane:
$\mathbb{H}_t \to \mathbb{H}$. To make the definition unique we
specify the behavior of $w_t(z)$ at infinity: \ba \la{asymptotic}
w_t(z) = z -\xi_t + \frac{2t}{z} + \ldots,\quad z\to\infty,
\end{align}
where $\xi_t$ is a real function of time uniquely determined by the
curve $\gamma_t$. The time-evolution of the map is given by the
Loewner equation \cite{Loewner-equation} \ba
\la{nonstochasticloewner} \der_t w_t(z) = \frac{2}{w_t(z)} -
\dot\xi_t,\quad w_t(z)\big|_{t=0}=z,\quad \xi_t\big|_{t=0}=0.
\end{align}
At each $z$ this equation is valid up to the time $t$ when
$w_t(z)=0$. When this happens the point $z$ no longer belongs to the
slit domain (it has been swallowed by the curve). For a simple curve
this happens when $z$ is hit by the curve: $\gamma_t=z$. Hence, the
Loewner function $w_t(z)$ maps the tip of the growing curve onto the
origin.

It is obvious that in order for $\gamma$ to be a continuous curve
without branching the forcing $\xi_t$ should be a continuous
function. Actually, certain types of forcing produce curves with
double points: $\gamma$ can self-touch. On the event of
self-touching a portion of the plane that has become encircled by
$\gamma$, is removed from $\mathbb{H}$ together with the $[0,t]$
portion of the curve, thus ensuring that the topology of
$\mathbb{H}_t$ does not change. The applicability range of the
Loewner equation, therefore, extends beyond the originally planned
description of one-dimensional objects, e.g. curves. In general, the
objects that grow according to the Loewner equation (the unions of
swallowed points) are called hulls.

The Loewner equation (\ref{nonstochasticloewner}) with a specified
forcing defines a curve or a hull in the upper half plane growing
with time, the process known as the chordal Loewner evolution. Its
basic property is that the final point of the evolution is
$\lim_{t\to\infty}\gamma_t=\infty$ i.e. a point on the boundary of
the domain $\mathbb{H}$. One can define the chordal Loewner
evolution in any domain $D$ topologically equivalent to $\mathbb{H}$
by a simple composition of maps. In this case by the Riemann theorem
there exists a conformal mapping $f(z):D\to\mathbb{H}$, and
$w_t(f(z))$ defines a growing curve $\gamma_t$ in the domain $D$: it
maps $D \backslash \gamma_t \to \mathbb{H}$. The final point of this
evolution $f^{-1}(\infty)$ necessarily lies on the boundary of $D$.

\subsection{Stochastic Loewner equation}

The Loewner evolution can be used to describe any curve which starts
from the origin and goes to infinity in the upper half-plane. We
would like to apply it to a conformally invariant curve produced in
CFT by the insertion of operators $\psi_{1,2}$ at $0$ and at
$\infty$. This dictates the necessary properties of the forcing
$\xi_t$.

Since $\gamma$ in a statistical system is a random curve the Loewner
forcing $\xi_t$ should be a stochastic process. This turns the
Loewner equation (\ref{nonstochasticloewner}) into a stochastic
differential equation \cite{Oksendal, Klebaner} which has the form
of a Langevin equation. In the rigorous sense it should be written
in terms of stochastic differentials because $\xi_t$ may (and will)
be nowhere-differentiable: \ba dw_t(z) &= w_{t+dt}(z) - w_t(z) =
\frac{2dt}{w_t(z)} - d\xi_t, & d\xi_t = \xi_{t+dt} - \xi_t.
\la{loewner-ito}
\end{align}
The r.h.s. is evaluated at the time $t$ (the It\^o convention for
stochastic analysis \cite{Oksendal}). This remark is important when
we deal with Markov processes: their behavior at any time is
completely independent of their earlier behavior. In the following
we will write stochastic differential equations with usual
time-derivatives, as in (\ref{nonstochasticloewner}), always
understanding them in the It\^o sense.

The properties that specify the driving function $\xi_t$ are
obtained from the following simple considerations. First of all, a
critical curve $\gamma$ in a translationally-invariant system has no
special preference of going to the left or to the right. The $\xi_t$
should have a distribution that is symmetric around zero leading to
the requirement \ba \la{symmetry} \xi_t = -\xi_t \quad \textrm{in
law}.
\end{align}
By this we mean the statistical identity: random quantities on
both sides of the equation are taken from the same distribution
but, in general, represent {\it different} realizations.

Next, the averaging over the ensemble of shapes of $\gamma$ can be
done in steps: first we can fix the shape of the curve up to a time
$t$ i.e. $\gamma[0,t]$, and average over the remaining part
$\gamma[t,\infty]$. In this averaging $\gamma[0,t]$ is seen as a
part of the domain boundary whose shape will be averaged later. With
a fixed $\gamma[0,t]$ the remaining part $\gamma[t,\infty]$ is a
conformally invariant curve in the domain $\mathbb{H}_t$. By the
conformal invariance we can map $\mathbb{H}_t$ onto $\mathbb{H}$ by
$w_t(z)$ and the image of $\gamma[t,\infty]$ should be statistically
identical to the original whole curve $\gamma[0,\infty]$ (see Fig.
\ref{slepicture}).
\begin{figure}[t]
\centering
\includegraphics[width=1\textwidth]{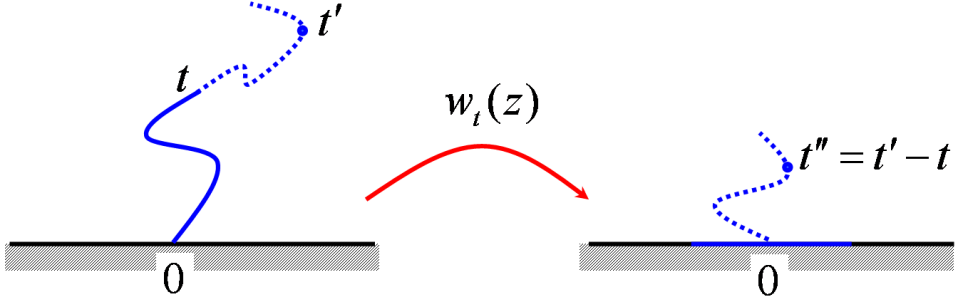}
\caption{The image of the $[t,t']$ part of the curve on the left
created by the mapping $w_t(z)$ is statistically identical to the
$[0,t'']$ part of the same curve on the right.} \la{slepicture}
\end{figure}
In particular, the image of any segment $\gamma[t,t']$ (here $t'>t$)
should be statistically identical to a segment $\gamma[0,t'']$ of
the original curve: \ba \gamma[0,t''] =
w_t(\gamma[t,t'])\quad\textrm{in law}.\nonumber
\end{align}
The fact that we can perform the averaging in steps means that the
r.h.s. and the l.h.s. are statistically independent. The application
of $w_{t''}$ maps the curve in the l.h.s. into the real axis, and
the same is done with the curve $\gamma[t,t']$ by the application of
$w_{t'}$. Therefore, $ w_{t'}(z)=w_{t''}(w_{t}(z))$ in law. At large
$z$ we can replace the Loewner maps with their asymptotic form
(\ref{asymptotic}) which allows us to deduce $t''=t'-t$ and arrive
to two equivalent statements \ba w_{t'}(z) &= w_{t'-t}(w_t(z)) \quad
\textrm{in law},
\la{sleinvariance} \\
\xi_{t'} &= \xi_t + \xi_{t'-t} \quad \textrm{in law}, \la{markov}
\end{align}
where the r.h.s. and the l.h.s. are statistically independent in
both formulas. Instead of eq. (\ref{sleinvariance}) we may equally
well write \ba w_{t'}(z)= w_t(w_{t'-t}(z))\quad\textrm{in law}.
\la{sleinvariance-1}
\end{align}

The only continuous stochastic process $\xi_t$ with stationary
independent increments as in eq. (\ref{markov}) and the symmetry
(\ref{symmetry}) is a driftless Brownian motion. Thus, it follows
from the conformal invariance and continuity of the curve that it
should be described by stochastic Loewner evolution (SLE). It is
defined by eq. (\ref{nonstochasticloewner}) where the forcing
$\xi_t$ is a Brownian motion. The requirement of continuity is
important here: it ensures that the curve does not branch and can be
chronologically ordered in a unique way. Hence, the probability
measure on a conformally invariant curve starting from the boundary
is related to the Brownian motion via the Loewner equation. There is
only one free parameter here: the strength of the noise $\kappa$
defined by \ba \ls \dot\xi_t\dot\xi_{t'}\rs =
\kappa\delta(t-t'),\nonumber
\end{align}
where we use the notation $\ls \ldots \rs$ for averages over
realizations of the Brownian motion $\xi_t$.

Different values of $\kappa$ must give rise to conformally invariant
curves in different statistical systems. Actually, the curve
$\gamma$ is simple i.e. never touches the boundary of the domain (or
itself, by conformal invariance), if and only if $\kappa \leqslant
4$ \cite{Rohde-Schramm} as in Fig. \ref{fig:dense dilute}. This is
consistent with parametrization of systems with dilute critical
curves by $\kappa < 4$ in Section \ref{sec:CFT}. At $\kappa>4$ the
Loewner transformation $w_t(z)$ maps the external perimeter of
$\gamma_t$ onto the real axis.

Next we turn to the formula (\ref{sleinvariance-1}) which we rewrite
using the sign $\circ$ for composition of maps as $w_t^{-1}\circ
w_{t'} = w_{t'-t}$ in law. Note that $w_t^{-1}\circ w_{t'}$ maps
$\mathbb{H}_{t'}$ onto $\mathbb{H}_t.$ Consider now an infinitesimal
time increment: $dt = t' - t$, for which we have $w_t^{-1} \circ
w_{t+dt} = w_{dt}$ in law. The infinitesimal Loewner map in the
r.h.s. is given by (see eq. (\ref{loewner-ito})) \ba w_{dt}(z) = z +
\frac{2dt}{z} - d\xi_t. \la{infw}
\end{align}
Then the statistically equivalent infinitesimal map $w_t^{-1}\circ
w_{t+dt}$ from ${\mathbb H}_{t+dt}$ to ${\mathbb H}_t$ is given by
the same equation: \ba w_t^{-1}(w_{t+dt}(z)) = z + \frac{2dt}{z} -
d\xi_t. \la{infwt}
\end{align}
We stress, however, that the equivalence of these maps is only
statistical and the Brownian motions $\xi_t$ in the two formulas
(\ref{infw}, \ref{infwt}) are two {\it different} realizations drawn
from the same statistical ensemble. Treating them as the same
function would lead to an inconsistency because no function $\xi_t$
(apart from a linear function $at$) satisfies the requirement
(\ref{markov}) in the exact, not statistical, sense.

\subsection{Martingales and correlation functions}

If a stochastic process $\mathcal{O}(t)$ is a martingale, its
average does not depend on time. Below we explain how a martingale
$\mathcal{O}(t)$ generated in SLE corresponds to a CFT holomorphic
operator $\mathcal{O}$ such that \cite{Bauer-Bernard-big-review}
\ba\la{nudela} \ls
\mathcal{O}(t)\rs=C_{\mathcal{O}}(\alpha_0)\frac{\langle
\psi_{1,2}(0)\mathcal{O}\psi_{1,2}(\infty)\Psi(\infty)\rangle_{\mathbb{H}}}
{\langle\psi_{1,2}(0)\psi_{1,2}(\infty)\rangle_{\mathbb{H}}}.\end{align}
The identification is up to a constant $C_{\mathcal{O}}$ which
depends on $\alpha_0$. We do not discuss it here.

Comparison with the two-step averaging (\ref{tralalala}) in which,
as we now know, the curve-creating operator is
$\mathcal{B}=\psi_{1,2}$, suggests the identification \ba
\mathcal{O}(t)\sim \langle
\mathcal{O}\Psi(\infty)\rangle_{\gamma_t}.\nonumber\end{align} We
will demonstrate the correspondence in the case of primary fields.

We will further explore the relation between CFT and SLE and show
that at $t\to\infty$ the SLE fields $\mathcal{O}(t)$ satisfy the
operator algebra of their counterparts in CFT. This is to be
expected. The size of $\gamma_t$ introduces a length scale $\sim
\sqrt t$. When it diverges at $t\to\infty$, the local fields in SLE
fluctuate in a scale-invariant way. In the following discussion we
consider only the dilute phase $\kappa\leqslant 4$.

\subsubsection{Primary fields in SLE}

The Loewner equation can be considered in any domain. The fields
related to the evolving curve transform under the conformal map $f$
which maps one domain onto another. We call an SLE primary field a
field which evolves under the Loewner equation and transforms as
$\mathcal{O}_h(z)\to f'^h(z) \mathcal{O}_h(f(z))$ under a change of
the domain. In particular, we can consider a pair of domains
$\mathbb{H}$ and $\mathbb{H}_t$ related by conformal map $w_t(z)$ in
which case \ba \mathcal{O}_h(z, t) = w_t'(z)^h \mathcal{O}_h(w_t(z),
0).\la{primaryfield-SLE}
\end{align}

Alternatively, we can use the pair of domains $\mathbb{H}_t$ and
$\mathbb{H}_{t+dt}$ related by an infinitesimal conformal
transformation $w_t^{-1}(w_{t+dt}(z))$ in which case \ba
\mathcal{O}_h(z, t+dt) = \big(\der_z w_t^{-1}(w_{t+dt}(z)) \big)^h
\mathcal{O}_h(w_t^{-1}(w_{t+dt}(z)), t).\nonumber
\end{align}
Now we use the stochastic nature of SLE: the statistical identity of
$w_t^{-1}\circ w_{t+dt}$ and $w_{dt}$ and the explicit expression
(\ref{infwt}) enables us to write \ba \mathcal{O}_h(z, t+dt) =
w_{dt}'(z)^h \mathcal{O}_h(w_{dt}(z), t) = \Big(1 - \frac{2dt}{z^2}
\Big)^h \mathcal{O}_h \Big(z + \frac{2dt}{z}-d\xi_t, t
\Big).\nonumber
\end{align}
Expanding the r.h.s. to the first order in $dt$ and to the second
order in $d\xi_t$ produces the evolution equation for
$\mathcal{O}_h(z,t)$: \ba \der_t \mathcal{O}_h(z, t) =
\Bigl(\frac{\kappa}{2} \mathcal{L}_{-1}^2 - 2\mathcal{L}_{-2} +
\dot\xi \mathcal{L}_{-1} \Bigr) \mathcal{O}_h(z, t),
\la{primaryfield}
\end{align}
where we used the notation (\ref{virasororep}) with $k=1$ and the
position of the curve-creating operator $x$ set to $0$. The
derivation is extended to a product of $k$ primary fields, when
operators $\mathcal{L}_{-1}$, $\mathcal{L}_{-2}$ act on $k$
variables.

The two ways of representing the stochastic evolution of
$\mathcal{O}_h(z, t)$ in the time-derivative of eq.
(\ref{primaryfield-SLE}) and in eq. (\ref{primaryfield}) are
equivalent but cannot be obtained from each other by a simple
substitution since they contain different realizations of Brownian
motion.

Since we understand stochastic equations in the It\^o sense, the
differential of the Brownian motion $\xi_{t+dt}-\xi_t$ in the last
term of eq. (\ref{primaryfield}) is multiplied by
$\der_z\mathcal{O}_h(z, t)$ evaluated at time $t$. Therefore, by the
Markov property of the Brownian motion this term vanishes upon
taking the average. The expectation value of a primary field then
satisfies the Feynman-Kac partial differential equation
\ba\la{multiprimaryfieldav} \der_t \ls \mathcal{O}_h(z,t) \rs =
\Bigl(\frac{\kappa}{2} \mathcal{L}_{-1}^2 - 2\mathcal{L}_{-2} \Bigr)
\ls \mathcal{O}_h(z,t) \rs,
\end{align}
and similarly for a product of $k$ fields. The evolution equation
for a primary field is obtained solely from its transformation law
under an infinitesimal conformal map. The same method allows to
obtain the evolution equation for any field with a specified
transformation law, not necessarily primary.

An example of a primary field is
\ba\la{simpleprimary}\mathcal{O}_h(z,t)=t^{\alpha_{1,2}\alpha_h}w_t'(z)^h,\end{align}
using the notation (\ref{Kac weights}) and (\ref{alphahdilute}). As
follows from the scaling law of SLE: \ba w_t(z)=
\frac{1}{a}w_{a^2t}(za),\quad \textrm{in law},\nonumber\end{align}
the solution of the equation (\ref{multiprimaryfieldav}) for a real
$z=x$ in this case must have the form
$t^{\alpha_{1,2}\alpha_h}F(x/\sqrt t)$ with the finite asymptote
$x^{2\alpha_{1,2}\alpha_h}$ at $t\to\infty$. In this limit the
fluctuations of $\mathcal{O}_h$ become scale-invariant and its
average corresponds to a correlation function in CFT. There local
fields have a singular short distance expansion. We now comment on
how it arises in the SLE approach.

Consider the product of two boundary fields $t^{\alpha_{1,2}
(\alpha_{h_1} + \alpha_{h_2} )} w'_t(x_1)^{h_1} w'_t(x_2)^{h_2}.$
Taking $x_1 = x_2$ we obtain $t^{\alpha_{1,2} (\alpha_{h_1} +
\alpha_{h_2} )} w'_t(x_1)^{h_1+h_2}, \la{ill-behaved}$ which
diverges at $t\to\infty$ due to convexity of $\alpha_h$. The
properly normalized field $t^{\alpha_{1,2} \alpha_{(h_1 + h_2)}}
w_t'(x_1)^{h_1+h_2}$ is different. This is an example of the short
distance singularity. It closely parallels the origin of such
singularities in field theory.

Since $\der_t\ls \mathcal{O}_h(z,t)\rs$ vanishes at $t\to\infty$,
the Feynman-Kac equation (\ref{multiprimaryfieldav}) in this limit
becomes identical to eq. (\ref{cfteq1}). Therefore, the SLE average
$\ls \mathcal{O}_h(z,\infty)\rs$ and the CFT correlation function
$\langle
\psi_{1,2}(0)V_{\alpha_h}(z)\Psi(\infty)\rangle_{\mathbb{H}}$ obey
the same equation. However, this fact alone does not allow to
identify them as the following simple remark shows. The CFT
correlation function is a holomorphic function
$z^{2\alpha_{1,2}\alpha_h}$ with a branch point at the origin. It
acquires a phase in going from a real $z$ to $-z$. However, the SLE
field $ t^{\alpha_{1,2}\alpha_h}w'_t(z)^h$ does not since $w'_t$ is
real on the real axis.

The proper choice of the primary field of weight $h$
is\footnote{\la{constants}This formula is understood up to a
multiplicative constant related to the constant
$C_{\mathcal{O}}(\alpha_0)$ in (\protect\ref{nudela}).}
\ba\la{vertexdef}
V^{\alpha_h}(z,t)=w_t(z)^{2\alpha_{1,2}\alpha_h}w'_t(z)^{h}.\end{align}
Its expectation value is holomorphic and can be identified with the
holomorphic CFT correlation function $\langle
\psi_{1,2}(0)V^{\alpha_h}(z)\Psi(\infty)\rangle=z^{2\alpha_{1,2}\alpha_h}$.
Indeed, its initial value is
$V^{\alpha_h}(z,t)\big|_{t=0}=z^{2\alpha_{1,2}\alpha_h}$ and as we
will show momentarily, its average does not depend on time.

Using the Loewner equation and the It\^o formula for a product of
two stochastic processes $X_t$ and $Y_t$: \ba\la{itoito}
d(X_tY_t)=X_tdY_t+Y_tdX_t+dX_tdY_t\end{align} with
$X_t=w_t(z)^{2\alpha_{1,2}\alpha_h}$ and $Y_t=w'_t(z)^h$ we find \ba
\der_tV^{\alpha_h}(z,t)=-\frac{2\alpha_{1,2}\alpha_h}{w_t(z)}V^{\alpha_h}(z,t)\dot\xi_t,
\nonumber\end{align} which yields $\der_t\ls V^{\alpha_h}(z,t)\rs=0$
i.e. $V^{\alpha_h}(z,t)=w_t(z)^{2\alpha_{1,2}\alpha_h}w'_t(z)^{h}$
is a martingale. We note that $\alpha_{1,2}=1/\sqrt\kappa$ and thus
$\alpha_{1,2}\xi_t$ is a Brownian motion $B_t$ with variance 1.

We conclude that $\ls V^{\alpha_h}(z,t)\rs$ satisfies the stationary
equation which is identical to (\ref{cfteq1}):\ba
\Bigl(\frac{\kappa}{4}\mathcal{L}_{-1}^2-\mathcal{L}_{-2}\Bigr)\ls
V^{\alpha_h}(z,t)\rs=0.\nonumber\end{align} at any time and not only
at $t\to\infty$. The same equation holds for a string of $k$ primary
fields.

In the remainder of this section we will further illustrate the
correspondence between SLE and Bose field.

\subsubsection{Bose field in SLE}

We now will discuss the SLE Bose field $\phi(z,t)$, a martingale
which corresponds to the holomorphic part of the Bose field
discussed in Sec. \ref{dilute phase}: \ba\phi(z,t)\sim\langle
\psi_{1,2}(0)\phi(z)\Psi(\infty)\rangle_{\gamma_t}.\nonumber\end{align}
It follows that $\phi(z,t)$ has the transformation law
(\ref{phi-transformation-dense}) and is a martingale. This fixes it
to be \cite{AC-geometry} \ba \phi(z,t) = i \sqrt2 \alpha_0 \log
w_t'(z) - \frac{i\mathcal{R}}{2}\log
w_t(z),\quad\mathcal{R}=\sqrt\frac{8}{\kappa}, \nonumber
\end{align} up to an additive constant (cf. footnote
\ref{constants}). Indeed, it follows from the Loewner equation that
$d\phi(z,t)=\frac{i\mathcal{R}}{w_t(z)}d\xi_t$ and therefore the
average of $\phi(z,t)$ equals its initial value
$\ls\phi(z,t)\rs=-\frac{i\mathcal{R}}{2} \log z$. The latter is just
the holomorphic part of the solution of the Laplace equation with a
step-like boundary condition $\varphi(x<0)=\pi\mathcal{R}$,
$\varphi(x>0)=0$.

The real field $\varphi(z,\bar z,t)=\phi(z,t)+\overline{\phi(z,t)}$
is compactified with radius $\mathcal{R}$. If we choose a point
$z_\gamma$ to lie on the slit $\gamma$ the value of $\varphi$ is
given by the direction $\vartheta$ of the tangential
vector\footnote{In fact, $\gamma$ is a fractal curve. It is
understood that $\vartheta$ is measured not on $\gamma$ but
nearby.}: \ba\varphi_{L}(z_\gamma)=2\sqrt2\alpha_0\vartheta
+\pi\mathcal{R},\quad
\varphi_{R}(z_\gamma)=2\sqrt2\alpha_0(\vartheta -\pi),
\la{windingangle}\end{align} where the additive constants correspond
to the two sides of the slit (Fig. \ref{fig: slefields}).
\begin{figure}[t]
\centering
\includegraphics[width=0.6\textwidth]{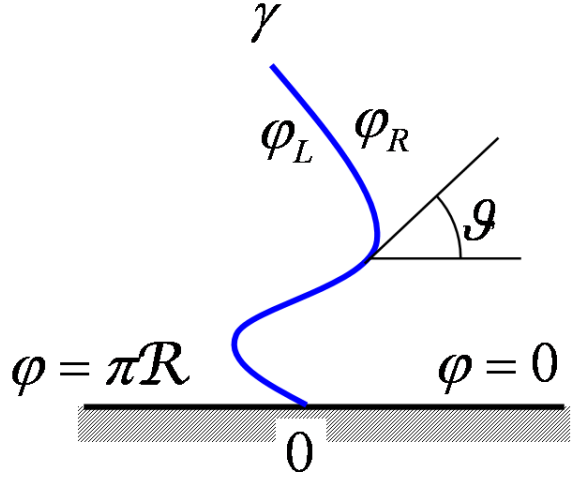}
\caption{The SLE field $\varphi$ on $\gamma$ measures the winding
angle $\vartheta$ (see eq. (\protect\ref{windingangle})). The
tangential component of the current $\partial_l\varphi$ on $\gamma$
measures the curvature.}\la{fig: slefields}
\end{figure}
The Bose field then measures the winding angle of $\gamma$. The
tangential vector of $\gamma$ is $e^{i\varphi_R/2\sqrt2\alpha_0},$
as in (\ref{flowlines}). The identification of the Bose field with
an angle emphasizes that in the dilute phase it is a pseudo-scalar.
The field fluctuates with $\gamma$ but its jump across $\gamma$ is
fixed:
\ba(\varphi_L-\varphi_R)\big|_{\gamma}=\pi(\mathcal{R}+2\sqrt2\alpha_0)=\sqrt{\frac{\kappa}{2}}\pi.\nonumber\end{align}
On the real axis $\varphi$ obeys the Dirichlet boundary condition:
$\varphi(x,t)=0$ for $x>0$ and $\varphi(x,t)=\pi\mathcal{R}$ for
$x<0$.

The case $\kappa=4$ (i.e. $\alpha_0=0$) is special: the values of
the field on $\gamma_t$ then are fixed to be $0$ and
$\pi\mathcal{R}$ as on the real axis.

The evolution equation for the Bose field follows from its
transformation law and the Loewner equation similarly to the
previous section. It is given by \ba \der_t \phi(z,t) =
\Bigl(\frac{\kappa}{2} \mathcal{L}_{-1}^2 - 2 \mathcal{L}_{-2} +
\dot\xi \mathcal{L}_{-1}\Bigr)\phi(z,t) -\frac{i
2\sqrt2\alpha_0}{z^2}, \la{boson}
\end{align}with $h=0$ in the definitions (\ref{virasororep}) of
$\mathcal{L}$'s. Since $\phi(z,t)$ is a martingale its average
$\ls\phi(z,t)\rs=-\frac{i\mathcal{R}}{2} \log z$ solves the equation
\ba\la{bosondecouple} \Bigl(\frac{\kappa}{4} \mathcal{L}_{-1}^2 -
\mathcal{L}_{-2} \Bigr) \ls\phi(z,t) \rs -\frac{i
\sqrt2\alpha_0}{z^2}=0.
\end{align}
This is the same equation as $\langle \psi_{1,2}(0) \phi(z)
\Psi(\infty) \rangle$ satisfies.

We now comment on how the operator algebra of the Bose field arises
in the SLE approach. With the help of the It\^o formula
(\ref{itoito}) we find from the Loewner equation the evolution of
the product of two Bose fields: \ba d(\phi(z_1,t) \phi(z_2,t)) =
d\phi(z_1,t)\phi(z_2,t) + \phi(z_1,t)d\phi(z_2,t) +
d\phi(z_1,t)d\phi(z_2,t).\nonumber
\end{align}
One can check that the last term (drift) is \ba d\phi(z_1,t)
d\phi(z_2,t) = - \frac{2 dt}{w_t(z_1)
w_t(z_2)}=d\log(w_t(z_1)-w_t(z_2)), \nonumber\end{align}and
therefore $\phi(z_1,t)\phi(z_2,t)$ is not a martingale. The
martingale reads \ba
\phi(z_1,t)\phi(z_2,t)-\log(w_t(z_1)-w_t(z_2)).\la{normalboson}\end{align}
This process is identified with the product of operators in CFT. The
subtracted term here is non-local. However, at $t\to\infty$ it
vanishes --- a manifestation of the fact that only in this limit
local fields of SLE fluctuate as local fields of CFT.

In the limit $z_1\to z_2$ the subtracted term becomes $\log (
{w_t(z_1)-w_t(z_2)})=\log w'_t(z_1)+\log(z_1-z_2)$. We define the
normal-ordered product by subtracting the logarithmic divergence
from (\ref{normalboson}): \ba \nol \phi^2(z,t)\nor =\phi^2(z,t)-\log
w'_t(z).\nonumber\end{align} The It\^o formula (\ref{itoito})
applied to a product of more than two bosons becomes the Wick
theorem. The normal-ordered exponential is found up to a
multiplicative constant as \ba \nol e^{i \sqrt2 \alpha_h \phi_t(z)}
\nor =w_t'(z)^{\alpha_h^2} e^{i \sqrt2 \alpha_h \phi_t(z)} = w_t^{2
\alpha_{1,2}\alpha_h} (z) w_t'(z)^h,\nonumber
\end{align}
which is just the definition (\ref{vertexdef}) of
$V^{\alpha_h}(z,t)$.

The holomorphic current of the Bose field is \ba j(z,t) =
\der\phi(z,t) = i \sqrt{2} \alpha_0 \frac{w_t''(z)}{w_t'(z)}
-i\frac{\mathcal{R}}{2}\frac{w_t'(z)}{w_t(z)}. \nonumber
\end{align}
Since $\varphi$ is a pseudo-scalar, the current is an axial vector.
On the curve $\gamma$ its tangential component is proportional to
the geodesic curvature of $\gamma$: \ba
j_l\big|_{\gamma}=2\sqrt2\alpha_0 K_\gamma.\nonumber\end{align} This
is consistent with the action (\ref{diluteaction}). At the tip of
the curve $\gamma_t$ the current has a vortex with a charge
proportional to $\alpha_0$.

The evolution of $j(z,t)$ and the equation for $\ls j(z,t)\rs$ are
found by differentiating eqq. (\ref{boson},\ref{bosondecouple}).
Differentiating (\ref{normalboson}) and taking $z_1\to z_2$ we get
the normal ordered square of the current \ba \nol j^2(z,t) \nor =
j^2(z,t) - \frac{1}{6}\{w_t,z\}, \nonumber
\end{align}
where $\{w,z\}$ is the Schwarz derivative: \ba \{w,z\} =
\frac{w'''(z)}{w'(z)}-\frac{3}{2}\Big(\frac{w''(z)}{w'(z)}\Big)^2.\nonumber
\end{align}

Finally, we can define the holomorphic stress-energy tensor in SLE
as a martingale with the proper transformation law: \ba z\to
f(z):\quad T(z,t) \to f'(z)^2T(f(z),t) + \frac{c}{12}
\{f,z\},\nonumber
\end{align}
with $c=1-24\alpha_0^2$. This fixes it to be \ba T(z,t) =-
\frac{1}{2} \nol \der\phi(z,t) \der\phi(z,t) \nor + i \sqrt2
\alpha_0 \der^2\phi(z,t),\nonumber
\end{align}which explicitly reads
\ba T(z,t)= \frac{c}{12}\{w_t,z\} +
h_{1,2}\left(\frac{w_t'(z)}{w_t(z)}\right)^2,\quad
h_{1,2}=\frac{6-\kappa}{2\kappa}.\nonumber\end{align} The
combination $T(z,t)dz^2$ can be used to determine the probability
that $\gamma$ intersects a segment of size $dz$ centered at $z$
\cite{Friedrich, Riva}.

The averages of the current and the stress-energy tensor are \ba \ls
j(z,t)\rs =-\frac{i\sqrt2\alpha_{1,2}}{z},\quad \ls
T(z,t)\rs=\frac{h_{1,2}}{z^2}\nonumber\end{align} We note that the
current and the stress-energy tensor measure the charge and the
weight of the curve-creating operator $\psi_{1,2}$.

We thus have seen that the stochastic Loewner evolution allows to
write the objects of field theory explicitly in terms of the shape
of the curve. These objects can be used to find distributions of
various characteristics of curves. As an application we compute in
the next section the harmonic measure of critical curves for
$\kappa\leqslant4$.

\section{Harmonic measure of critical curves}
\la{sec:harmonic measure} In this section we apply the results of
the previous sections to compute the multifractal spectrum of the
harmonic measure of dilute ($\kappa\leqslant 4$) critical curves. In
the dense phase $\kappa>4$ these results apply to external
perimeters of critical curves.

Harmonic measure quantifies geometry of complicated plane domains
\cite{harmonic measure-1, harmonic measure-2}. We start with basic
definitions.

\subsection{Harmonic measure of critical curves}
\la{subsec:definition harmonic measure}

Consider a closed simple curve $\gamma$. One can imagine that it is
made of a conducting material and carries a total electric charge
one. The harmonic measure of any part of $\gamma$ is defined as the
electric charge of this part. In what follows we will pick a point
of interest $z_0$ on the curve and consider a disc of a small radius
$r\ll R$ centered at $z_0$. It surrounds a small part of $\gamma$
and we define $\mu(z_0, r)$ to be the harmonic measure of this part
i.e. the electric charge in it.

Then we consider the moments \ba M_h = \sum_{i=1}^N \mu(z_i,
r)^h,\la{moments}
\end{align}
where $h$ is a real power and $N$ is the number of discs needed to
cover $\gamma$ as in Fig. \ref{fig:multifractality}. As the radius
$r$ gets smaller and the number of discs $N$ gets larger, these
moments scale as \ba M_h \sim \Bigl(\frac{r}{L}\Bigr)^{\tau(h)},
\quad \frac{r}{L}\rightarrow 0, \nonumber
\end{align}
where $L$ is the typical size of the curve.

\begin{figure}[t]
\centering
\includegraphics[width=0.6\textwidth]{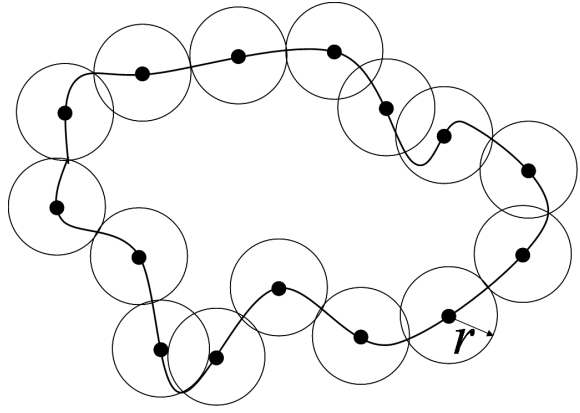}
\caption{A curve covered by discs.} \la{fig:multifractality}
\end{figure}

The function $\tau(h)$ is called the multifractal spectrum of the
curve $\gamma$. This function encodes a lot of information of the
curve $\gamma$. It also has some simple properties. First of all,
since all $0 < \mu(z_i, r) \leqslant 1$, the moments $M_h$ are well
defined for any real $h$ and the function $\tau(h)$ is
non-decreasing: $\tau(h) \leqslant \tau(h')$ for any $h < h'$.
Secondly, if $h=1$ the sum in (\ref{moments}) is equal to the total
charge of the cluster and therefore does not scale with $r$,
producing the normalization condition $\tau(1) = 0$. Thirdly, if we
set $h=0$, $M_0$ is simply the number $N$ of discs of radius $r$
necessary to cover the curve $\gamma$ so that by definition the
fractal (Hausdorff) dimension of $\gamma$ is $d_f = -\tau(0)$.

If the curve $\gamma$ were smooth we would have a simple relation
$\tau(h) = h-1$. For a fractal curve one defines the anomalous
exponents $\delta(h)$ by \ba \tau(h) &= h - 1 + \delta(h).\nonumber
\end{align}
Also, the generalized multifractal dimensions of a fractal curve
$\gamma$ are defined as $D(h) = \tau(h)/(h-1)$ (so that $D(0) =
d_f$). A theorem \cite{Makarov-1, Makarov-2} states that $ D(1) =
\tau'(1) = 1.$

If the curve $\gamma$ is random, so are its moments $M_h$, and we
can average them over the ensemble of curves. It is natural to
assume that the summation over the points $z_i$ in eq.
(\ref{moments}) is equivalent to the ensemble average. Hence we can
write \ba \ls M_h\rs = N \ls \mu(r)^h \rs \sim
\Bigl(\frac{r}{L}\Bigr)^{\tau(0)} \ls \mu(r)^h \rs,\nonumber
\end{align}
where now the harmonic measure $\mu(r)$ is measured at any point $z$
on $\gamma$ and $\ls\ldots\rs$ indicates the ensemble averaging. We
define the local multifractal exponent $\w\tau(h)=\tau(h)-\tau(0)$
at a point $z$ by \ba \ls {\mu(r)^h} \rs &\sim
\Bigl(\frac{r}{L}\Bigr)^{\w\tau(h)}. \la{local exponent}
\end{align}

The ensemble of critical curves is suitable for multifractal
analysis. There are a few generalizations of the simple closed curve
considered above. First of all, the curve $\gamma$ need not be
closed or stay away from the system boundaries. If $\gamma$ touches
a boundary we can supplement it with its mirror image across this
boundary. The definition of $\mu(z,r)$ can be naturally extended to
the cases when $z$ is the endpoint of $n$ critical curves on the
boundary or in the bulk. If $n$ is even, the latter case can be also
seen as $n/2$ critical curves passing through $z$. In particular,
$n=2$ corresponds to the situation of a single curve in the bulk
considered above.

When $z$ is the endpoint of $n$ critical curves on the boundary or
in the bulk we define the corresponding scaling exponents similarly
to eq. (\ref{local exponent}): \ba \ls {\mu(z, r)^h} \rs &\sim r^{h
+ \Delta^{(n)}(h)}, & \ls {\mu(z, r)^h} \rs &\sim r^{h +
\Delta^{(n)}_{\mathrm{bulk}}(h)}. \la{exponent definition}
\end{align}
In the case of a single curve we will drop the superscript, for
example, $\Delta(h)\equiv \Delta^{(1)}(h)$.

These exponents were first obtained by means of quantum gravity in
\cite{Duplantier-PRL, duplantier}. For a critical system
parameterized by $\kappa$ the results read \ba \Delta(h) &=
\frac{\kappa-4 + \sqrt{(\kappa-4)^2 + 16\kappa h}} {2\kappa} =
\frac{\sqrt{1-c+24h} - \sqrt{1-c}} {\sqrt{25-c} - \sqrt{1-c}},
\la{Delta(h)-again}
\\
\Delta^{(n)}(h) &= n \Delta(h), \la{simple boundary exponent}
\\
\Delta^{(n)}_{\mathrm{bulk}}(h) & = - \frac{h}{2} +
\Big(\frac{1}{16} + \frac{n-1}{4\kappa}\Big) \big(\kappa - 4 +
\sqrt{(\kappa - 4)^2 + 16\kappa h}\big). \la{simple bulk exponent}
\end{align}
Remarkably, $\Delta(h)$ is the gravitationally dressed dimension $h$
as given by the KPZ formula of 2D quantum gravity
\cite{Knizhnik:1988ak}. Connections of this formula to SLE were
discussed in \cite{bauer-bernard-2}. Starting in the next section we
will show how to obtain these exponents formulating the $c\leqslant
1$ CFT via the Bose field, where they are also written in a more
transparent way.

\subsection{Harmonic measure and uniformizing conformal maps}

Let us consider a conformal map $w(z)$ of the exterior of $\gamma$
to a standard domain. Usually we will choose the upper half plane
but sometimes the exterior of a unit circle is more convenient. We
choose the map so that the point of interest $z$ on the curve is
mapped onto itself, choose it to be the origin $z=0$ and normalize
the conformal map so that  $w(\infty)=\infty$ and
$w'(\infty)=\mathrm{const.}$ In the case of the upper half plane
we choose $w'(\infty)=1$ as we did in SLE (see Section
\ref{sec:SLE}), while for the unit circle $w'(\infty)=\rho^{-1}$
where $\rho$ is the conformal radius of $\gamma$.
\begin{figure}[t]
\centering
\includegraphics[width=0.8\textwidth]{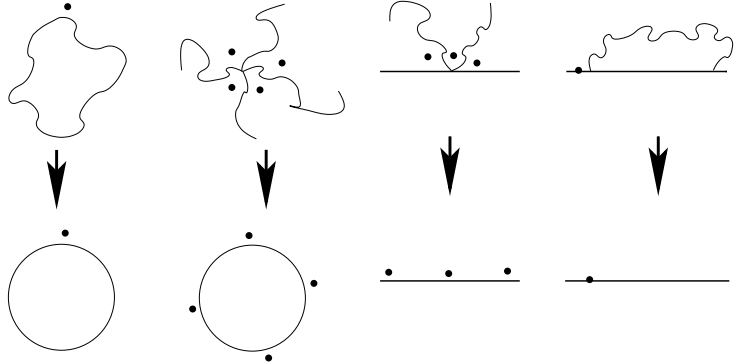}
\caption{The uniformizing conformal maps for various cases
considered. The dots denote that points where the electric field is
measured.} \la{fig: mappings}
\end{figure}

The scaling of $w(z)$ near the origin is directly related to that of
the harmonic measure. Indeed, since $\mu(0,r)$ is the charge inside
the disc of radius $r$ by the Gauss theorem it is equal to the flux
of the electric field through the boundary of this disc. This, in
turn, should scale as the circumference of the disc times a typical
absolute value of the electric field at a distance $r$ from the
origin i.e. $|w'(r)|$. This leads to a scaling relation \ba \mu(0,r)
\sim r|w'(r)|,\nonumber
\end{align}
which allows to rewrite the definitions (\ref{exponent definition})
as \ba \ls {|w'(r)|^h} \rs &\sim r^{\Delta^{(n)}(h)}, & \ls
{|w'(r)|^h} \rs &\sim r^{\Delta^{(n)}_{\mathrm{bulk}}(h)}. \nonumber
\end{align}

The relation between the scaling of the harmonic measure and the
derivative of a uniformizing map allows further generalizations.
Namely, we can measure the electric field in more than one point.
Consider $n$ critical curves emanating from the origin. Close to the
origin the curves divide the plane into $n$ sectors if the origin
lies in the bulk and $n + 1$ sectors if it lies on the boundary.
Then we can study objects like \ba & \ls \prod_i|w'(z_i)|^{h_i}
\rs,\nonumber
\end{align}
where no two $z_i$'s lie in the same sector. The case when the
electric field is not measured in some sectors is done by setting
$h_i=0$ in them. We will see how to express these quantities as CFT
correlation functions. In the case when $z_i$'s are all at the
distance $r$ from the origin these averages scale as
$r^{\Delta^{(n)}(h_1,\ldots h_{n+1})}$ and
$r^{\Delta^{(n)}_{\mathrm{bulk}}(h_1,\ldots h_n)}$ with the higher
multifractal exponents related to the multifractal exponents
(\ref{Delta(h)-again} -- \ref{simple bulk exponent}) as
\cite{duplantier} \ba \Delta^{(n)}(h_1,\ldots h_{n+1}) & =
\sum_{i=1}^{n+1} \Delta^{(n)}(h_i) + \frac{\kappa}{2}
\sum_{i<j}^{n+1} \Delta(h_i) \Delta(h_j),
\la{higher boundary exponent} \\
\Delta^{(n)}_{\mathrm{bulk}}(h_1,\ldots h_n) & = \sum_{i=1}^n
\Delta_{\mathrm{bulk}}^{(n)}(h_i) + \frac{\kappa}{4} \sum_{i<j}^n
\Delta(h_i) \Delta(h_j). \la{higher bulk exponent}
\end{align}
In the case when $z_i$'s are independent variables the answer is
more complicated. However, in the case of a single curve at the
boundary it reads: \ba \ls |w'(x_1)|^{h_1}|w'(x_2)|^{h_2} \rs \sim
|x_1|^{\Delta(h_1)}|x_2|^{\Delta(h_2)}|x_1-x_2|^{\Delta(h_1,h_2)-
\Delta(h_1)-\Delta(h_2)}. \la{boundaryoppositetwopoint}
\end{align}

\subsection{Calculation of multifractal spectrum}

Here we obtain formulas (\ref{Delta(h)-again} -- \ref{simple bulk
exponent}, \ref{higher boundary exponent}, \ref{higher bulk
exponent}). We will consider in detail the simplest case of harmonic
measure of a single curve near the boundary. The other cases contain
a few technical differences. The calculation is based on the
two-step averaging of Sec. \ref{twostepaverage}.
\subsubsection{Boundary multifractal exponent from CFT}
All averages over the ensemble of critical curves are some
correlation functions of the Bose field. Averages of $w'^h$ are
related to correlation functions of primary fields $\mathcal{O}_h$
in the presence of curve-creating operators. We start with the case
of a single curve starting from the boundary, when this operator is
$\psi_{1,2}$ and find the boundary multifractal exponent
$\Delta(h)$.

We use the two-step averaging result (\ref{tralalala}) where we
write $\mathcal{B}\equiv\psi_{1,2}$ and choose $\mathcal{O}$ to be a
boundary vertex operator
$\mathcal{O}_h(r)=e^{i\sqrt2\alpha_h\phi(r)}$. We uniformize the
exterior of $\gamma$ by mapping it to the upper half plane. Being a
primary operator of weight $h$, $\mathcal{O}_h(r)$ transforms as
$\mathcal{O}_h \to w'(r)^h \mathcal{O}_h(w(r))$ while $\Psi(\infty)$
does not transform because of the normalization of $w(z)$ at
infinity. The transformation relates the correlation function in the
exterior of $\gamma$ to a correlation function in the upper half
plane: \ba \langle
\mathcal{O}_h(r)\Psi(\infty)\rangle_\gamma=w'(r)^h \langle
\mathcal{O}_h(w(r))\Psi(\infty)\rangle_\mathbb{H }. \la{2pt
correlator}
\end{align}
Note that the correlation function in the r.h.s. has no singular
dependence on $r$, hence, as long as we stay in the limit $r\ll |L|$
it is a constant prefactor. We then obtain a scaling
relation\footnote{We drop the absolute value since on the real axis
$w'$ is positive.} \ba \ls {w'(r)^h} \rs \sim \langle
\mathcal{O}_h(r) \psi_{1,2}(0)
\Psi(\infty)\rangle_\mathbb{H}.\nonumber
\end{align}
The primary field $\Psi(\infty)$ should be chosen such as to satisfy
the physical condition $\Delta(0)=0$. This formula establishes the
relation between the multifractal spectrum of critical curves and
the correlation functions. Explicitly, the correlation function
reads $\langle
e^{i\sqrt2\alpha_h\phi(r)}e^{i\sqrt2\alpha_{1,2}\phi(0)}
e^{i\sqrt2(2\alpha_0-\alpha_h-\alpha_{1,2})\phi(\infty)}\rangle$,
where $0,r,\infty$ are three points on the real axis. It remains to
find how it scales with $r$.

To find the $r$-dependence of this correlation function, we use the
fusion of vertex operators (\ref{fusionrule}) and obtain the result
(\ref{Delta(h)-again}) written in a compact and suggestive form: \ba
\la{eb} \Delta(h) = 2 \alpha_{1,2} \alpha_h,\quad
\alpha_h=\alpha_0+\sqrt{\alpha_0^2+h}.
\end{align}

An immediate generalization to the statistics of the harmonic
measure of $n$ curves reaching the system boundary at the same point
is obtained by replacing $\psi_{1,2} \to \psi_{1, n + 1}$. Since
$\alpha_{1, n + 1} = -n\alpha_-/2 = n \alpha_{1,2}$, this
immediately leads to \ba \Delta^{(n)}(h) = 2 \alpha_{1, n + 1}
\alpha_h = n \Delta(h),\nonumber
\end{align}
which is the same as eq. (\ref{simple boundary exponent}).

We also remark that nothing compels us to measure the electric field
on the real axis. Instead of the boundary field $\mathcal{O}_h(r)$
we could take a bulk primary field $\mathcal{O}_{h'}(z, \bar z)$
where $|z|=r$. The weight $h'$ should be chosen such that when the
holomorphic part $\mathcal{O}_{h'}(z)$ is fused with its image
$\mathcal{O}_{h'}(\bar z)$ the boundary field $\mathcal{O}_h(\frac{z
+ \bar z}{2})$ with weight $h$ is recovered.
\subsubsection{Boundary multifractal exponent from SLE}
For comparison, we compute the multifractal spectrum of a single
curve in the SLE approach (see e.g. \cite{Lawler-book}).

In SLE we must compute $\ls w_t'(r)^h\rs$ in the limit $t\to\infty$
since the size of the curve $L\sim \sqrt t$. With the choice of the
primary field as in (\ref{simpleprimary}) the eq.
(\ref{multiprimaryfieldav}) at $t\to\infty$ reads \ba
\Bigl(\frac{\kappa}{2}\der_r^2+\frac{2}{r}\der_r -
\frac{2h}{r^2}\Bigr)\ls\, w'^h_t(r)\rs=0.\nonumber\end{align} The
solution is a power law $\ls\, w_t'(r)^h\rs\sim
r^{2\alpha_{1,2}\alpha_h}$, in agreement with (\ref{eb}).

\subsubsection{Calculation of higher boundary multifractal exponents}

We consider $n$ non-intersecting critical curves growing from the
origin on the boundary (the real axis). It will be convenient to
assume that the curves end somewhere in the bulk thus forming a
boundary star (e.g. the third picture in Fig. \ref{fig: mappings}).
Let $w(z)$ be the conformal map of the slit domain formed by the
star to the upper half plane normalized at infinity.

We want to find the scaling of the average \ba \ls {|w'(z_1)|^{h_1}
\ldots |w'(z_{n+1})|^{h_{n+1}}} \rs,\nonumber
\end{align}
where $z_i$'s are all close to the origin, no two of them lying in
the same sector. The latter condition will be automatically
satisfied in the subsequent calculation due to the following: if in
a particular realization some two points $z_i$ and $z_j$ happen to
be in the same sector, then $w(z_i) - w(z_j) \to 0$ as the size of
the star $L\to\infty$.

Since $n$ curves starting from the origin on the boundary are
produced by the operator $\psi_{1,n+1}(0)$ we now consider a
boundary CFT correlation function with several ``probes'' of the
harmonic measure: \ba C = \big\langle \prod_{i=1}^{n+1}
\mathcal{O}_{h_i'}(z_i, \bar z_i)
\psi_{1,n+1}(0)\Psi(\infty)\big\rangle_{\mathbb{H}}.\nonumber
\end{align}
This correlation function is found as \ba C\sim\prod_i
|z_i|^{4\alpha_{1,n+1} \alpha_i'} \prod_{i<j} |z_i - z_j|^{4
\alpha_i' \alpha_j'} \prod_{i,j} (z_i - \bar z_j)^{2 \alpha_i'
\alpha_j'}.\end{align} When all $z_i$ are at the same distance $r$
from the origin it scales as \ba C \sim r^{4 \alpha_{1,n+1} \sum_i
\alpha_i' + 8 \sum_{i<j} \alpha_i' \alpha_j' + 2 \sum_i
\alpha_i'{}^2}. \la{C-scaling-1}
\end{align}
As before, this correlation function is equal to the statistical
average of a certain correlation function in the fluctuating domain,
and we further apply the uniformizing map $w(z)$ to transform this
domain into the upper half plane $\mathbb{H}$: \ba C = \,\, \ls
{\prod_i |w'(z_i)|^{2h_i'} \big\langle \prod_i
\mathcal{O}_{h_i'}\big(w(z_i), \overline{w(z_i)}\big)
\Psi(\infty)\big\rangle_{\mathbb{H}}} \rs. \la{C-2}
\end{align}
Unlike eq. (\ref{2pt correlator}), the correlation function inside
the average scales as: \ba\prod_i
(w(z_i)\!-\!\overline{w(z_i)})^{2\alpha_i'{}^2} \prod_{i<j} |w(z_i)
\!-\! w(z_j)|^{4 \alpha_i' \alpha_j'} \prod_{i \neq j} \big(w(z_i)
\!-\! \overline{w(z_j)}\big)^{2 \alpha_i'
\alpha_j'},\la{sameforboth}
\end{align}
where $\alpha'_i\equiv
\alpha_{h'_{i}}=\alpha_0+\sqrt{\alpha_0^2+h'_i}$. We specifically
separated the $i=j$ terms since only they contribute to the
short-distance behavior. All the other terms insure that the
realizations of the curves in which any two points $z_i$ end up in
the same sector are suppressed since the distances $w(z_i) - w(z_j)$
are then small. Then the short-distance dependence of eq.
(\ref{C-2}) is \ba C \sim\, \ls {\prod_i |w'(z_i)|^{2h_i'}
(w(z_i)-\overline{w(z_i)})^{2\alpha_i'{}^2}} \rs.\nonumber
\end{align}
Insofar as the scaling with $r$ is concerned, we further approximate
$w(z_i)-\overline{w(z_i)} \sim |z_i| |w'(z_i)| \sim r |w'(z_i)|$.
This gives \ba C \sim r^{2\sum_i \alpha_i'{}^2} \ls {\prod_i
|w'(z_i)|^{2h_i' + 2\alpha_i'{}^2}} \rs. \la{C-scaling-2}
\end{align}
We denote the exponents inside the average \ba h_i = 2h_i' +
2\alpha_i'{}^2.\la{differentweights}
\end{align}These are the weights of operators whose conformal charges
$\alpha_{h_i}=\alpha_0+\sqrt{\alpha_0^2+h_i}$ are just $2\alpha'_i$.

Finally, comparing eqs. (\ref{C-scaling-1}) and (\ref{C-scaling-2}),
we get the result \ba \ls {|w'(z_1)|^{h_1} \ldots
|w'(z_{n+1})|^{h_{n+1}}} \rs \propto r^{\Delta^{(n)}(h_1,\ldots
h_{n+1})},\nonumber
\end{align}
with the higher multifractal exponent \ba
\Delta^{(n)}(h_1,...h_{n+1}) &=
2\alpha_{1,n+1}\sum_{i=1}^{n+1}\alpha_{h_i} +
2\sum_{i<j}^{n+1}\alpha_{h_i} \alpha_{h_j},\nonumber
\end{align}
which is the formula (\ref{higher boundary exponent}).

If $z_i$'s are kept independent the computation becomes more
complicated. It can be done to the end in the case of a single
curve. In this case the points can be taken on the boundary and $C$
reduces to a four-point function which contains $\psi_{1,2}$:
\ba\la{firstformula}C(x_1,x_2)\sim\prec\!\!
w'(x_1)^{h_1}w'(x_2)^{h_2}\!\!\succ\,\, \sim \langle
\mathcal{O}_{h_1}(x_1)\mathcal{O}_{h_2}(x_2)\psi_{1,2}(0)\Psi(\infty)\rangle_{\mathbb{H}},
\end{align}
where $x_1$ and $x_2$ are two points on the real axis. Since
$\psi_{1,2}$ is degenerate on level 2, $C$ satisfies the equation
(\ref{cfteq1}). Same result can be obtained within the SLE approach.
Up to a normalization the solution is \cite{cft} \ba
\la{solution}C\sim
x_1^{\Delta(h_1)}x_2^{\Delta(h_2)}(x_1-x_2)^{\Delta -
\Delta(h_1)-\Delta(h_2)} F\Big(\frac{x_1}{x_1-x_2}\Big),
\end{align}
\ba \Delta = h_\infty - h_{1,2} - h_1 - h_2,\nonumber
\end{align}where $h_\infty$ is the weight of $\Psi(\infty)$
and $F(z)$ satisfies a hypergeometric equation: \ba &
\frac{\kappa}{4} z(1-z) F''(z) + \Bigl[1 + \frac{\kappa}{2}
\Delta(h_1) - \Bigl( 2 + \frac{\kappa}{2} \Delta(h_1) +
\frac{\kappa}{2} \Delta(h_2) \Bigr) z \Bigr]
F'(z) \nonumber \\
& - (\Delta(h_1,h_2) - \Delta) F(z) = 0. \nonumber
\end{align}
One of the two possible solutions of this equation diverges at small
$z$ and the other one is regular. If $x_2 \to \infty$ the dependence
of the correlation function on $x_1$ should be $C(x_1,\infty)\sim
\langle\mathcal{O}_{h_1}\psi_{1,2}\Psi(\infty)\rangle\sim
x_1^{\Delta(h_1)}$. We therefore should pick the regular solution of
the hypergeometric equation. Then for $C$ to be symmetric with
respect to $(x_1,h_1)\leftrightarrow (x_2,h_2)$, $F(z)$ should be a
constant. A constant is a solution of the hypergeometric equation
only if the weight at infinity is fixed by setting \ba\Delta =
\Delta(h_1,h_2).\nonumber
\end{align}
We then obtain the formula (\ref{boundaryoppositetwopoint}).

We note that the function $C(x_1, x_2)$ as defined in eq.
(\ref{firstformula}) is real and positive for any $x_1$, $x_2$.
Therefore, the constant $F$ must have such a phase as to make the
expression (\ref{solution}) positive.
\subsubsection{Calculation of bulk multifractal exponents}

Calculation of the bulk multifractal behavior is done in much the
same way as on the boundary, so we go straight to the general case
of higher bulk exponents. The method of SLE is not yet developed for
this case.

Let the critical system occupy the whole complex plane and be
conditioned to have $n$ critical curves growing from a single point
in which we place the origin $z=0$. We will assume that $z=0$ is the
only common point of these curves, since the local results around
this point are unaffected by the curves' behavior at large
distances. The curves then form a star. We define the conformal map
$w(z)$ of the slit domain formed by the star to the exterior of a
unit circle (the second picture in Fig. \ref{fig: mappings}) and
normalize $w(z)$ at infinity as before: $w(\infty)=\infty$,
$w'(\infty)=\rho^{-1}$, where $\rho$ is the conformal radius of the
star.

Close to the origin the curves divide the plane into $n$ sectors. We
consider a quantity \ba \ls {|w'(z_1)|^{h_1}\ldots |w'(z_n)|^{h_n}}
\rs,\nonumber
\end{align}
where $z_i$'s are points close to the origin, no two of them lying
in one sector. As before, this is not a serious constraint because
if some two points $z_i$ and $z_j$ happen to be in the same sector,
$w(z_i)-w(z_j)\rightarrow 0$ when the size of the star $L\to\infty$.

Since $n$ curves starting from the origin in the bulk are produced
by the operator $\psi_{0,n/2}(0),$ we consider the correlation
function \ba C_{\text{bulk}} = \big\langle \prod_{i=1}^{n}
\mathcal{O}_{h_i'}(z_i, \bar z_i)
\psi_{0,n/2}(0)\Psi(\infty)\big\rangle, \nonumber
\end{align}
where $h'_i$ are related to $h_i$ as in (\ref{differentweights}):
$h_i$ have twice larger holomorphic charges:\ba 2\alpha'_i =
\alpha_{h_i}=\alpha_0+\sqrt{\alpha_0^2+h_i}.\nonumber
\end{align}
The correlation function is found to be \ba C_{\text{bulk}} \sim
\prod_i |z_i|^{4\alpha_{0,n/2} \alpha_i'} \prod_{i<j} |z_i - z_j|^{4
\alpha_i' \alpha_j'} \sim r^{4\alpha_{0,n/2} \sum_i \alpha_i' + 4
\sum_{i<j} \alpha_i' \alpha_j'},\nonumber
\end{align}assuming that all $z_i$ are of the order of $r$.
Proceeding exactly as in the boundary case we use the two-step
averagin to rewrite the correlation function $C_{\text{bulk}}$
through the conformal map of the fluctuating star: \ba
C_{\text{bulk}} = \,\, \ls {\prod_i |w'(z_i)|^{2h_i'} \big\langle
\prod_i \mathcal{O}_{h_i'}\big(w(z_i), \overline{w(z_i)}\big)
\Psi(\infty)\big\rangle_{{\mathbb C}\setminus \mathrm{star}}} \rs.
\nonumber
\end{align}
The correlation function inside the average is found to be the same
as (\ref{sameforboth}) although $w(z)$ is defined differently.
Combining the results we obtain \ba \ls {|w'(z_1)|^{h_1}\ldots
|w'(z_n)|^{h_n}} \rs \sim r^{\Delta_{\mathrm{bulk}}^{(n)}(h_1,\ldots
h_n)}, \nonumber
\end{align}
with the higher bulk exponent \ba
\Delta_{\mathrm{bulk}}^{(n)}(h_1,\ldots h_n) = \sum_{i=1}^n
\Delta_{\mathrm{bulk}}^{(n)}(h_i)+\sum_{i<j}^
{n}\alpha_{h_i}\alpha_{h_j}, \nonumber
\end{align}
where \ba \Delta_{\mathrm{bulk}}^{(n)}(h) = 2\alpha_{0,n/2} \alpha_h
- \frac{1}{2} \alpha_h^2 = (2\alpha_{0,n/2}-\alpha_0)\alpha_h -
\frac{h}{2} \nonumber
\end{align}
is the scaling exponent of a single $\ls {|w'(z)|^h} \rs$ in the
presence of $n$ critical curves in the bulk. These are the results
quoted in eqs. (\ref{simple bulk exponent}, \ref{higher bulk
exponent}).

\section{Conclusions}

In this work we attempted to clarify the subtle relation between
the Gaussian Bose field description of critical behavior of
statistical models and stochastic geometry of critical curves
occurring in these systems. We further studied the relation
between two complementary approaches: the algebraic approach of
conformal field theory and a more direct approach of stochastic
\mbox{(Schramm-)} Loewner evolution.

A natural extension of this work is a study of global properties of
critical curves such as the distribution of the area and harmonic
moments of a closed critical loop.

The critical curves considered in this paper all appear at 2D
critical points described by CFTs with central charges $c \leqslant
1$. There are other CFTs which in addition to conformal symmetry
have other symmetries, e.g. current algebras. These theories allow
central charges $c > 1$.

We have recently extended the SLE approach to the Wess-Zumino model
with the current algebra SU$(2)_k$ \cite{BGLW}. In this case the SLE
trace carries an additional spin degree of freedom. The analog of
the Gaussian field takes values in the Lie group. The critical
curves which appear in such a system can also be studied by the
methods described here.

\section*{Acknowledgements}

We have benefitted from discussions with D.Bernard and M.Bauer. We
thank the Kavli Institute for Theoretical Physics at UCSB where this
paper was finished.

This research was supported in part by the National Science
Foundation under Grant No. PHY99-07949, the NSF MRSEC Program under
DMR-0213745, DMR-0540811, the NSF Career award DMR-0448820, the
Sloan Research Fellowship from Alfred P. Sloan Foundation, and the
Research Innovation Award from Research Corporation.


\begin{thebibliography}{99}

\bibitem{polyakov} A.~M.~Polyakov, JETP Lett. {\bf 12}, 381 (1970).

\bibitem{Belavin:1984vu}
A.~A.~Belavin, A.~M.~Polyakov, and A.~B.~Zamolodchikov, Nucl.\
Phys.\ B {\bf 241}, 333 (1984).

\bibitem{cft}
P. Di Francesco, P. Mathieu, and D. Senechal, {\it Conformal field
theory} (Springer, 1999).

\bibitem{Duplantier-PRL}
B. Duplantier, Phys. Rev. Lett. {\bf 84}, 1363 (2000); arXiv:
cond-mat/9908314.

\bibitem{duplantier}
B. Duplantier, in {\it Fractal geometry and applications}, Proc.
Sympos. Pure Math. {\bf 72}, part 2, Amer. Math. Soc., 2004;
arXiv: math-ph/0303034.

\bibitem{BRGW-harmonic-measure-PRL}
E. Bettelheim, I. Rushkin, I. A. Gruzberg, and P. Wiegmann, Phys.
Rev. Lett. {\bf 95}, 170602 (2005); arXiv: hep-th/0507115.

\bibitem{Knizhnik:1988ak}
V.~G.~Knizhnik, A.~M.~Polyakov, and A.~B.~Zamolodchikov, Mod.\
Phys.\ Lett.\ A {\bf 3}, 819 (1988).

\bibitem{Schramm}
O. Schramm, Israel J. Math. {\bf 118}, 221 (2000); arXiv:
math.PR/9904022.

\bibitem{Bauerbernard}
M. Bauer and D. Bernard, in {\it String theory: from gauge
interactions to cosmology}, 41, NATO Sci. Ser. II Math. Phys. Chem.,
{\bf 208}, Springer, Dordrecht, (2006); arXiv: cond-mat/0412372.

\bibitem{Bauer-Bernard-big-review}
M. Bauer and D. Bernard, Phys. Rep. {\bf 432}, 115 (2006); arXiv:
math-ph/0602049.

\bibitem{Cardy2}
J. Cardy, Ann. Phys. {\bf 318}, 81 (2005); arXiv:
cond-mat/0503313.

\bibitem{Gruzberg-review}
I. A. Gruzberg, J. Phys. A {\bf 89}, 12601 (2006); arXiv:
math-ph/0607046.

\bibitem{Kager-Nienhuis}
W. Kager and B. Nienhuis, J. Stat. Phys. {\bf 115}, 1149 (2004);
arXiv: math-ph/0312056.

\bibitem{Lawler-book}
G. F. Lawler, {\it Conformally invariant processes in the plane}.
Mathematical Surveys and Monographs, 114. American Mathematical
Society, Providence, RI, 2005.

\bibitem{Lawler2}
G. F. Lawler, in {\it Random walks and geometry}, 261, Walter de
Gruyter GmbH \& Co. KG, Berlin 2004; available online at URL
\verb|http://www.math.cornell.edu/~lawler/papers.html|

\bibitem{Schramm-review}
O. Schramm, in {\it Current developments in mathematics}, 2000, 233,
Int. Press, Somerville, MA, 2001.

\bibitem{Schramm-overview}
O. Schramm, to appear in the ICM 2006 Madrid Proceedings, arXiv:
math.PR/0602151.

\bibitem{Werner1}
W. Werner, in {\it Lectures on probability theory and statistics}.
Lecture Notes in Mathematics, {\bf 1840}. Springer-Verlag, Berlin,
2004; arXiv: math.PR/0303354.

\bibitem{Werner2}
W. Werner,  Probab. Surv. {\bf 2}, 145 (2005) (electronic); arXiv:
math.PR/0307353.

\bibitem{nienhuis}
B. Nienhuis, in {\it Phase transitions and critical phenomena},
edited by C. Domb (Academic Press, 1987), vol. 11.

\bibitem{cardyoperators}
J. L. Cardy, Nucl. Phys. {\textbf{B240}}, 514 (1984);
{\textbf{B324}}, 581 (1989).

\bibitem{recommended}
M. Bauer, D. Bernard and J. Houdayer, J. Stat. Mech. (2005) P03001;
arXiv: math-ph/0411038.

\bibitem{Schulze}
J. Schulze, Nucl.\ Phys.\ {\bf B489}, 580 (1997); arXiv:
hep-th/9602177.

\bibitem{kawai}
S.~Kawai, J.\ Phys.\ A {\bf 36}, 6547 (2003); arXiv: hep-th/0210032.

\bibitem{loop models 1}
H. Saleur, Phys. Rep. {\bf 184}, 177 (1989).

\bibitem{loop models 2}
B. Duplantier, Phys. Rep. {\bf 184}, 229 (1989).

\bibitem{loop models 3}
C. Vanderzande, {\it Lattice models of polymers}, Cambridge
University Press, 1998.

\bibitem{kondev 1}
J. Kondev, Int. J. Mod. Phys. {\bf B11}, 153 (1997); arXiv:
cond-mat/9607181.

\bibitem{kondev 2}
J. Kondev, Phys. Rev. Lett. {\bf 78}, 4320 (1997); arXiv:
cond-mat/9703113.

\bibitem{SS-gaussian-field}
O. Schramm and S. Sheffield, arxiv: math.PR/0605337.

\bibitem{Rohde-Schramm}
S. Rohde and O. Schramm, Ann. of Math. (2) {\bf 161}, 883 (2005);
arXiv: math.PR/0106036.

\bibitem{df}
Vl. S. Dotsenko, V. A. Fateev, Nucl. Phys. {\textbf{B240}}, 312
(1984).

\bibitem{BB-zigzag}
M. Bauer and D. Bernard, in Proceedings of the NATO conference {\it
``Conformal invariance and random spatial processes''}, Edinburgh
(2003); arXiv: math-ph/0401019.

\bibitem{bauer-bernard-1}
M. Bauer and D. Bernard, Phys. Lett. B {\bf 543}, 135 (2002) ;
arXiv: math-ph/0206028.

\bibitem{bauer-bernard-2}
M. Bauer and D. Bernard, Commun. Math. Phys. {\bf 239}, 493
(2003); arXiv: hep-th/0210015.

\bibitem{bauer-bernard-5}
M. Bauer and D. Bernard, Phys. Lett. B {\bf 583}, 324 (2004);
arXiv: math-ph/0310032.

\bibitem{kadanoff}
L. P. Kadanoff, J. Phys. A {\bf 11}, 1399 (1978).

\bibitem{polchinski}
J. Dai, R.G. Leigh, and J. Polchinski, Mod. Phys. Lett. A {\bf 4},
2073 (1989).

\bibitem{AC-geometry}
O. Schramm and S. Sheffield, unpublished; see Sheffield's lectures
at \verb|http://www.fields.utoronto.ca/audio/05-06/#percolation_SLE|

\bibitem{Loewner-equation}
L. V. Ahlfors, {\it Conformal invariants: topics in geometric
function theory}, McGraw-Hill, New York (1973).

\bibitem{Oksendal}
B. \O ksendal, {\it Stochastic differential equations},
Springer-Verlag, Berlin, 2003.

\bibitem{Klebaner}
F. C. Klebaner, {\it Introduction to stochastic calculus with
applications}, Imperial College Press, London, 1998.

\bibitem{Friedrich}
R. Friedrich and W. Werner, Comm. Math. Phys. {\bf 243}, 105 (2003);
arXiv: math-ph/0301018.

\bibitem{Riva}
B. Doyon, V. Riva, and J. Cardy, Comm. Math. Phys. {\bf 268}, 687
(2006); arXiv: math-ph/0511054.

\bibitem{harmonic measure-1}
J. B. Garnett and D. E. Marshall, {\it Harmonic measure},
Cambridge University Press, 2005.

\bibitem{harmonic measure-2}
C. Pommerenke, {\it Boundary behaviour of conformal maps}, Springer,
1992.

\bibitem{Makarov-1}
N. G. Makarov, Proc. London Math. Soc. {\bf 51}, 369 (1985).

\bibitem{Makarov-2}
N. G. Makarov, St. Petersburg Math. J. {\bf 10}, 217 (1999).

\bibitem{BGLW}
E. Bettelheim, I. A. Gruzberg, A. W. W. Ludwig, and P. Wiegmann,
Phys. Rev. Lett. {\bf 95}, 251601 (2005); arXiv: hep-th/0503013.


\end{thebibliography}
\end{document}